\documentclass[10pt,letterpaper]{article}
\usepackage{opex3}
\usepackage{subfigure}
\usepackage{cite}

\newcommand{\V}[1]{\mathbf{#1}}      
\newcommand{\M}[1]{\mathbf{#1}}          
\newcommand{\T}{^{\rm T}}             
\newcommand{\Abs}[1]{\left\vert #1\right\vert}

\begin{document}

\title{Beam shaping for laser-based adaptive optics in astronomy}

\author{Cl\'ementine~B\'echet,$^{1*}$ Andr\'es~Guesalaga,$^1$ Benoit~Neichel,$^{2~3}$ Vincent~Fesquet,$^2$ H\'ector~Gonz\'alez-N\'u\~nez,$^1$ Sebasti\'an~Z\'u\~niga,$^4$ Pedro~Escarate,$^4$ and Dani~Guzman$^1$}

\address{$~^1$ Pontificia Universidad Cat\'olica de Chile, 4860 Vicu\~na Mackenna, Santiago, Chile\\
  $~^2$Gemini Observatory, c/o AURA, Casilla 603, La Serena, Chile\\
  $~^3$Aix Marseille Universit\'e, CNRS LAM, UMR 7326, 13388 Marseille, France\\
  $~^4$ Universidad T\'ecnica Federico Santa Mar\'ia, Avenida Espa\~na
  1680, Valpara\'iso, Chile}

\email{$^*$CBechetP@ing.puc.cl} 



\begin{abstract}
  The availability and performance of laser-based adaptive optics (AO)
  systems are strongly dependent on the power and quality of the laser
  beam before being projected to the sky. Frequent and time-consuming
  alignment procedures are usually required in the laser systems with
  free-space optics to optimize the beam. Despite these procedures,
  significant distortions of the laser beam have been observed during
  the first two years of operation of the Gemini South multi-conjugate
  adaptive optics system (GeMS).

  A beam shaping concept with two deformable mirrors is investigated
  in order to provide automated optimization of the laser quality for
  astronomical AO. This study aims at demonstrating the
  correction of quasi-static aberrations of the laser, in both
  amplitude and phase, testing a prototype of this two-deformable
  mirror concept on GeMS. The paper presents the results of the
  preparatory study before the experimental phase. An algorithm to
  control amplitude and phase correction, based on phase retrieval
  techniques, is presented with a novel unwrapping method. Its
  performance is assessed via numerical simulations, using aberrations
  measured at GeMS as reference. The results predict effective
  amplitude and phase correction of the laser distortions with about
  120 actuators per mirror and a separation of 1.4~m between the
  mirrors. The spot size is estimated to be reduced by up to 15\%
  thanks to the correction. In terms of AO noise level, this has the
  same benefit as increasing the photon flux by 40\%.
\end{abstract}





\section{The need for laser beam optimization}
\label{sec:GemsContext}

The performance of adaptive optics (AO) is strongly dependent on the
quality of the wavefront sensing measurements from the guide stars
because they directly affect the correction error
\cite{Rousset1999a}. For the Shack-Hartmann (SH) wavefront sensors,
currently the better adapted sensor for laser guide stars on large
telescopes, the standard deviation measurement error ($\sigma_{\rm
  {meas}}$) is a function of the spot size and the number of photons
received per subaperture and per frame according to
\cite{Rousset1999a}:
\begin{equation}
  \label{eq:sigmaE}
  \sigma_{\rm {meas}} \propto \frac {\theta_{\rm {image}}} {\sqrt{N_{\rm ph}}}\,,
\end{equation}
where $\theta_{\rm {image}}$ is the full width at half maximum (FWHM)
of the spot image in the focal plane of a SH subaperture, and $N_{\rm
  ph}$ is the average number of photons received in a SH subaperture
each frame. Therefore, the smaller the spot image or the greater the
number of received photons, the better the measurement accuracy. This
article analyzes the case of the Gemini South multi-conjugate AO
system (GeMS) \cite{RigautNeichel2014a} which uses a single sodium
laser to generate a constellation of 5 laser guide stars on sky. The
size of the spot image $\theta_{\rm {image}}$ depends on various
parameters, among which the laser spot size in the sodium layer, the
elongation of the seen spot due to the sodium layer thickness, the
diffraction produced by the SH subaperture size and the atmospheric
seeing. 

For the first two years of operations of GeMS during the season of low
sodium concentration (December-January), the system frame rate had to
be frequently reduced compared to its maximum of 800~Hz in order to
guarantee the average detection of at least $135$ photons per
subaperture per frame from each laser guide star
\cite{NeichelRigaut2014a,DOrgevilleDiggs2012a}. This minimum flux is
required for adequate closed-loop AO correction by GeMS. From
Eq.~(\ref{eq:sigmaE}), if the image spot size $\theta_{\rm {image}}$
is reduced by 15\%, it is equivalent for a given AO noise level to
increase the number of photons by 40\%. Laser-based AO systems usually being
designed with little margin in terms of expected photons return, it
appears all the more important to minimize the spot size.

In laser specifications, the smallest laser spot possible is obtained
if the 589~nm beam to be projected has a beam quality factor $M^2=1$
\cite{Siegman1990a}. Any discrepancy from this ideal shape results in
$M^2$ values greater than 1. An estimation of the $M^2$ of a laser can
be computed using a SH to measure the beam irradiance and wavefront at
waist \cite{NealAlford1996a} and by using the discrete Fourier
transform of the resulting complex field, the far-field distribution
can also be computed. From each of these fields, the standard
deviation of the $x$ and $y$ axis margin distribution of the beam
intensity, can be obtained. Using the notations $D_x$ and $\theta_x$
for this quantity along $x-$direction at the waist and in the
far-field respectively, the $M^2$ is estimated by
\begin{equation}
  \label{eq:M2-estim}
  M^2_x = \frac{\pi D_x \theta_x} {4^3 \lambda} 
\end{equation}

\begin{figure}[!ht]
\centering
\includegraphics[width=0.35\linewidth]{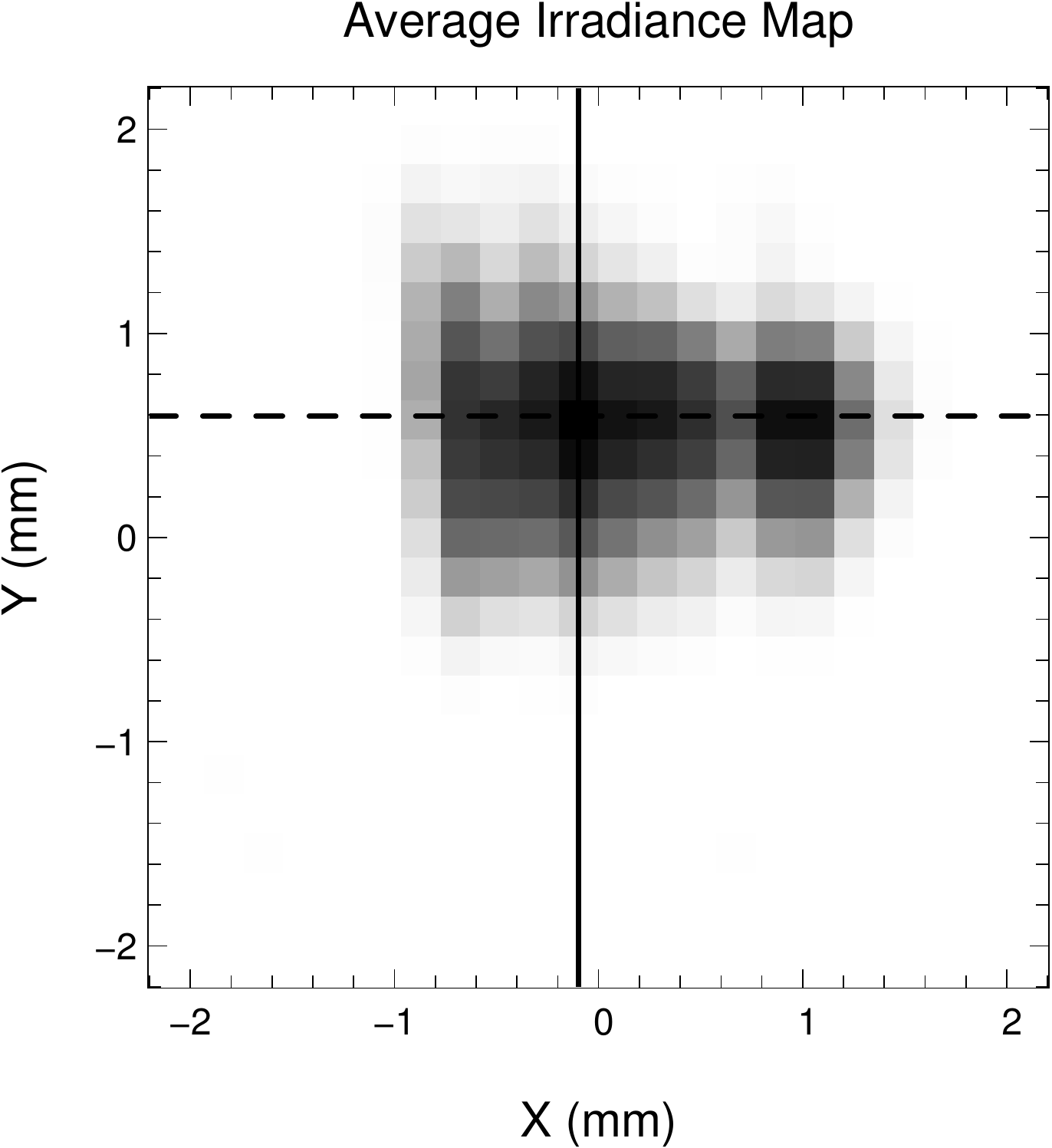}
\includegraphics[width=0.35\linewidth]{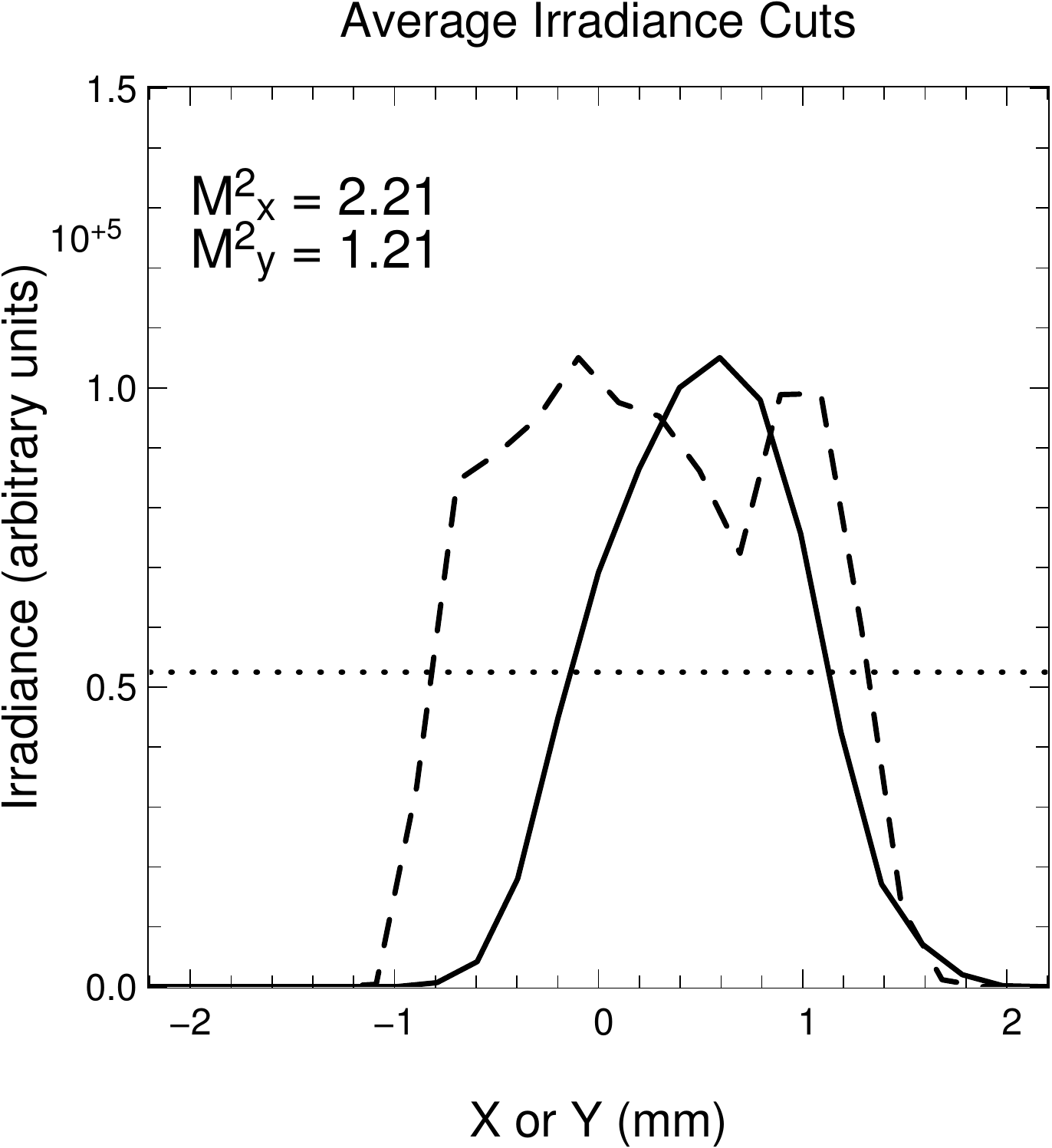}
\includegraphics[width=0.35\linewidth]{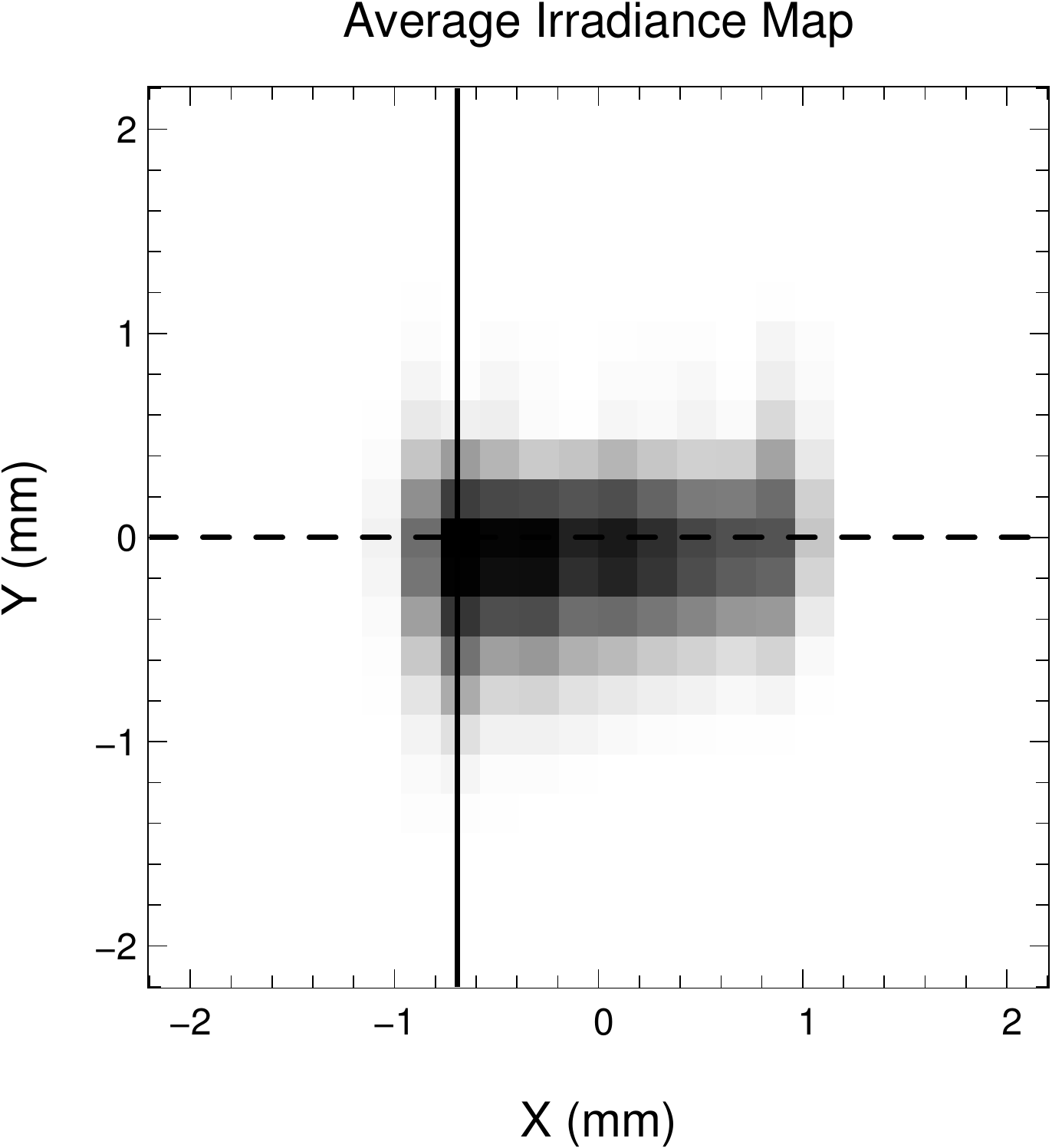}
\includegraphics[width=0.35\linewidth]{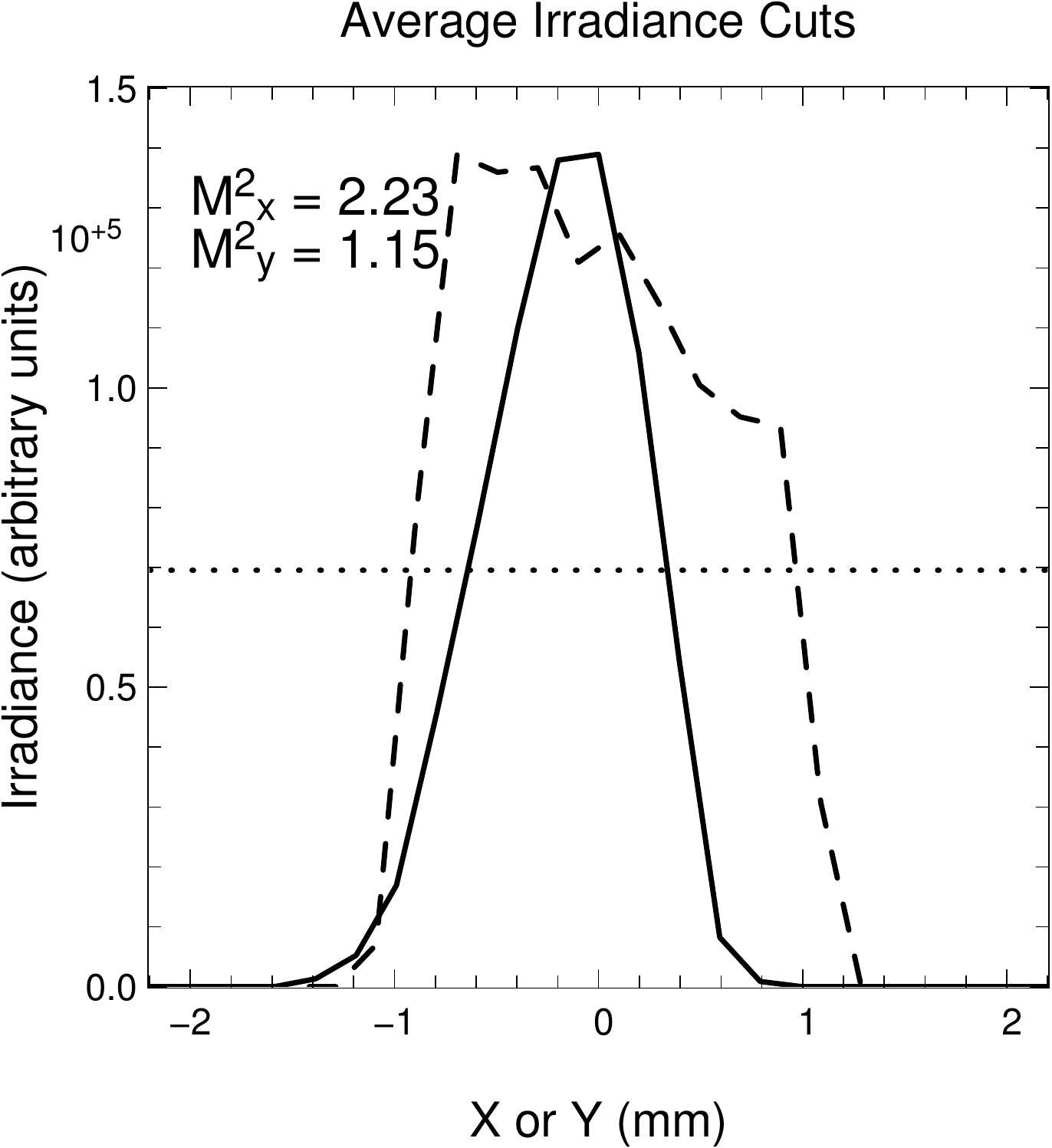}
\caption{Average irradiance map over 20 frames obtained from a $37
  \times 37$ Shack-Hartmann sensor located at the output of the laser,
  during a run in April 2013 (top) and a run in October 2013
  (bottom). {\bf Left:} Irradiance map, with orthogonal lines of cuts
  through the maximum. {\bf Right:} Cuts along $x$ (dashed) and
  $y$-axis (solid) of the irradiance map above. The dotted horizontal
  line marks the half of the maximum.}
\label{fig:irrmap}       
\end{figure}

During the 2011/2012 commissioning campaign, the measured $M^2$
factors on GeMS were below 1.5 and 1.3, in $x$ and $y$ respectively
\cite{DOrgevilleDiggs2012a}. A beam quality $M^2$ close to 1 has not
been achieved so far with GeMS laser. The $M^2$ values even worsened
during the first half of 2013, revealing significant degradation of
the laser beam quality with time. The laser beam has been measured
right at the output of the laser bench using a dedicated
Shack-Hartmann sensor. The obtained irradiance maps are shown in
Fig.~\ref{fig:irrmap} for sets of measurements taken in April and
October 2013. Note that these measurements were both made after
several days of alignment optimization, \textit{i.e.} on the best
laser shape at each time. From the data of April 2013, $M^2$ factors
of 2.21 and 1.21 along $x$ and $y$ directions respectively were
found. In October 2013, the corresponding $M^2$ factors were 2.23 and
1.15.

These values highlight the strong distortion of the amplitude along
$x$-axis, also visible in all plots of Fig.~\ref{fig:irrmap}. The
dotted line represents the half of the maximum to quantify the FWHM
and it shows that the beam diameter is between 1.5 and 2 times larger
along $x$ than along $y$. The beam quality is always worse in the
$x$-axis with GeMS because it corresponds to the unguided dimension of
the amplification stages in the waveguide amplifier module
\cite{DOrgevilleDiggs2012a}. In addition, a slow increase of the $M^2$
values over the length of the runs has also been noticed, suggesting
that the system suffers from constant misalignment
\cite{DOrgevilleDiggs2012a}.

The impact of this degraded shape of the beam on the laser intrinsic
spot size has been previously quantified by including the measured
irradiance of April 2013 (top of Fig.~\ref{fig:irrmap}) in simulations
of the uplink propagation of the laser from the launching telescope to
the mesosphere \cite{BechetGuesalaga2013a}. No phase aberrations were
considered, but the beam was simulated for amplitude shapes different
from that of a perfect Gaussian shape. The same uplink simulation code
is used to evaluate the impact of amplitude and phase distortions on
the short-exposure spot size at the sodium layer. Previous on-sky
measurement of the laser intrinsic spot size of $1.0"$ has been
reported for GeMS \cite{dOrgevilleDaruich2008a}. This number cannot be
directly compared to the values obtained with the simulation because
there is not a strictly linear dependence of the laser return flux and
the short-exposure irradiance focused at the sodium altitude. However,
GeMS laser is not expected to induce saturation at the sodium layer,
so the variance of the short-exposure irradiance over the SH
subaperture field-of-view provides a good estimate of the spot size at
a first order \cite{HolzlohnerBonaccini-Calia2008a}. In addition, the
vertical distribution of the sodium layer is not taken into account in
this computation, but its effect could be considered to globally
increase of the spot size by 10\%
\cite{HolzlohnerBonaccini-Calia2008a}. Simulating a seeing of
0.73~arcseconds at 500~nm (median seeing over 33 GeMS observations
nights between December 2012 and June 2013 \cite{NeichelRigaut2014a}),
on averaged over 100 successive independent atmospheres, the obtained
FWHM of the short-exposure spot (arcseconds) is:
\begin{itemize}
\item $0.61"$ for a perfect Gaussian beam with FWHM of 25~cm.
\item $0.69"$ and $0.71"$ considering only amplitude aberrations
  measured in April 2013 and in October 2013 respectively.
\item $0.74"$ considering both amplitude and phase aberrations of the
  laser beam. The same value $0.74"$ is obtained for April 2013 and for
  October 2013.
\end{itemize}
For a given AO noise level (in Eq.~(\ref{eq:sigmaE})), the reduced
size of the short-exposure laser spot with a perfect Gaussian beam
compared to the distorted beam would be equivalent to an increase of
the received number of photons by 30 to 45\%.

The beam degradation is understood as the result of quasi-static
aberrations in both the laser bench and the beam transfer optics (BTO)
path \cite{DOrgevilleDiggs2012a}. First, the co-alignment of the
optical system for sum frequency generation used in GeMS is highly
subject to temperature hysteresis, as well as telescope
vibrations. This critical alignment tends to drift along a
run. Secondly, the BTO is in a closed environment, with no turbulence,
but the outside temperature can vary by 10 to 15 degrees, so
aberrations along the BTO are subject to slow evolutions.  Frequent
and constraining calibrations and alignment procedures (quasi-static
aberrations) are thus required in both parts to optimize the beam to
launch. These complex and time-consuming alignment procedures
currently used will strongly reduce the availability of GeMS
\cite{NeichelRigaut2014a}. It may also affect the future generations
of AO requiring high power laser and using beam transport by mirrors,
as planned for the Thirty Meter Telescope \cite{Ellerbroek2013a}. It
is worth noting that recent research and development of fiber lasers
for sodium guide stars \cite{FengTaylor2009a} have shown very
promising laboratory results to produce lasers with better $M^2$
quality and high power. There is still few on-sky experience to
guarantee that this new technology can solve all the problems for an
AO system like GeMS, requiring a 50-Watt laser. Therefore the study of an
automatic technique to improve the beam shape of the current laser in
GeMS is considered highly desirable.

The concept of a beam shaping system for laser-based astronomical AO
was recently proposed \cite{GuesalagaNeichel2012a} and its potential
benefit on the GeMS signal-to-noise ratio has been highlighted with
preliminary simulations of the uplink propagation of the laser toward
the sodium layer \cite{BechetGuesalaga2013a}. This concept is studied
here in greater detail, quantifying with numerical simulations the
expected performance of the method with respect to design
parameters. These results will be further used to develop an
experimental prototype for such beam correction to be tested on GeMS
laser.

In Sect.~\ref{sec:FieldConjugation}, the architecture with two
deformable mirrors for amplitude and phase correction is
detailed. In Sect.~\ref{sec:Eqs}, the iterative algorithm used to
deduce the commands for both mirrors is described with emphasis given
to the iterative unwrapper required to solve the problem in
practice. Last in Sect.~\ref{sec:simu-field-conj}, preliminary
simulations of the beam shaping correction method are presented. The
results confirm the ability to correct both amplitude and phase with
this type of architectures. The influence of parameters like the
separation distance between the mirrors, the number of iterations of
the algorithm and the number of actuators of the mirrors are
emphasized.

\section{2-DM concept for laser beam shaping}
\label{sec:FieldConjugation}

In order to optimize the laser beam to be launched, correction of both
amplitude and phase distortions of the beam is required. This is
referred to as optical field conjugation or amplitude-phase correction
\cite{KanevLukin1991a}.

The presented concept takes its origins in the beam shaping research
for laser radar and active coherent imaging (\textit{e.g.}
\cite{Frieden1965a,Ogland1978a,Shafer1982a,RhodesShealy1980a,EismannTai1989a}). In
these first studies, the common goal was to produce an optical device
which modifies the beam coming from a given source to lead to a
desired irradiance distribution in the near-field. In order to avoid
loss of energy, only refractive or reflective surfaces must be used to
redistribute the energy. Early works were focused on static shaping
and lead to the difficult manufacturing of very specific optical
components \cite{Ogland1978a,Shafer1982a,RhodesShealy1980a}. Eismann
\textit{et al.}  \cite{EismannTai1989a} proposed a more versatile
solution where they transform a Gaussian beam into a uniformly
illuminated field using two computer-generated holographic
elements. During the past two decades, several wavefront correction
technologies were developed to match this low loss of energy and to
provide also adaptive correction for evolving distortions introduced
by propagation through turbulent media. Among
these, 
the micro-electro-mechanical systems (MEMS) deformable mirrors (DM)
\cite{BakerStappaerts2004a} are now compact and robust device for beam
shaping at least cost. In our study, we foresee to use two MEMS
DMs.

\begin{figure}[htb!]
  \centering
  \mbox{\subfigure[]{\includegraphics[width=0.5\linewidth]{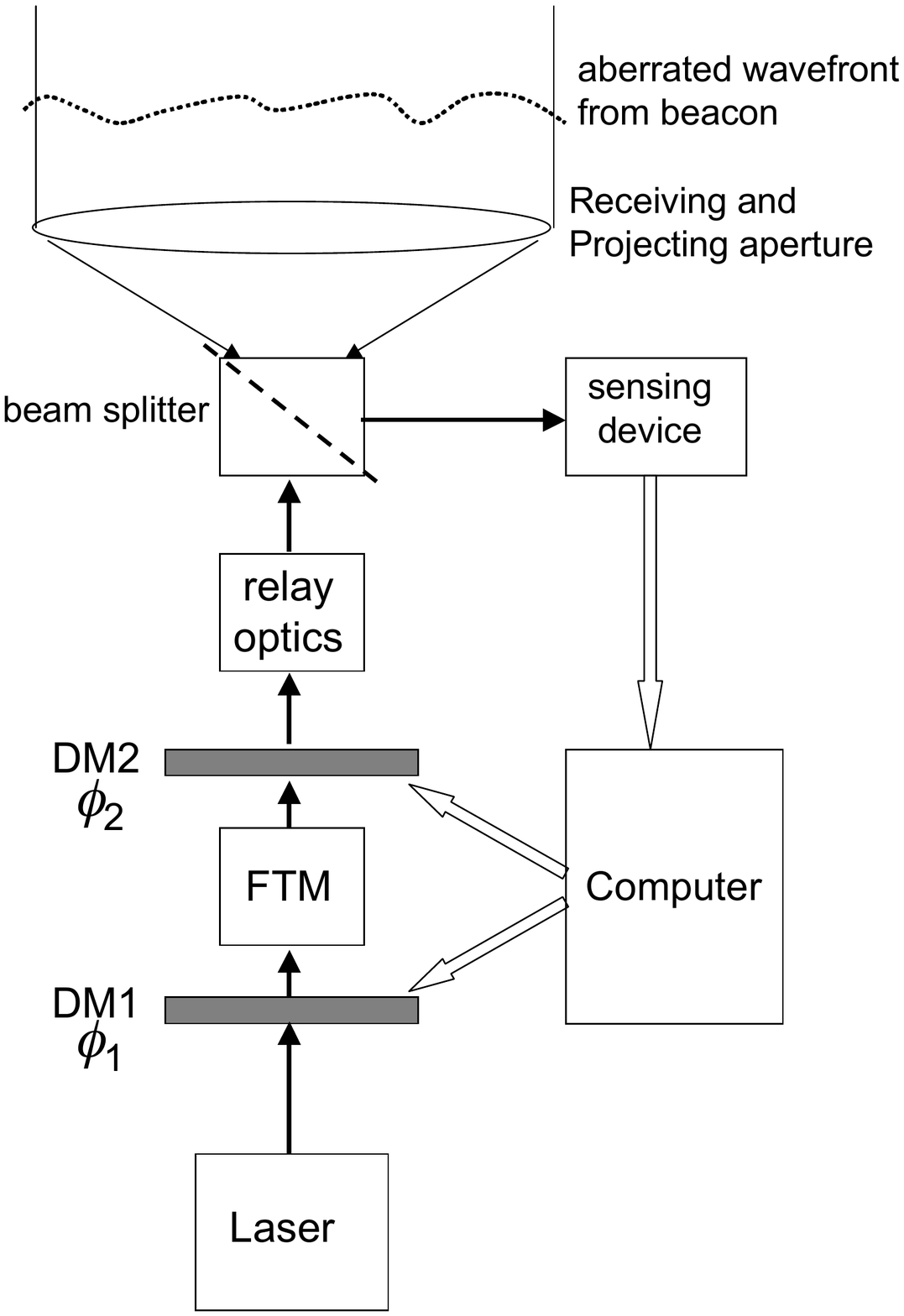}} \vline
  \subfigure[]{\includegraphics[width=0.5\linewidth]{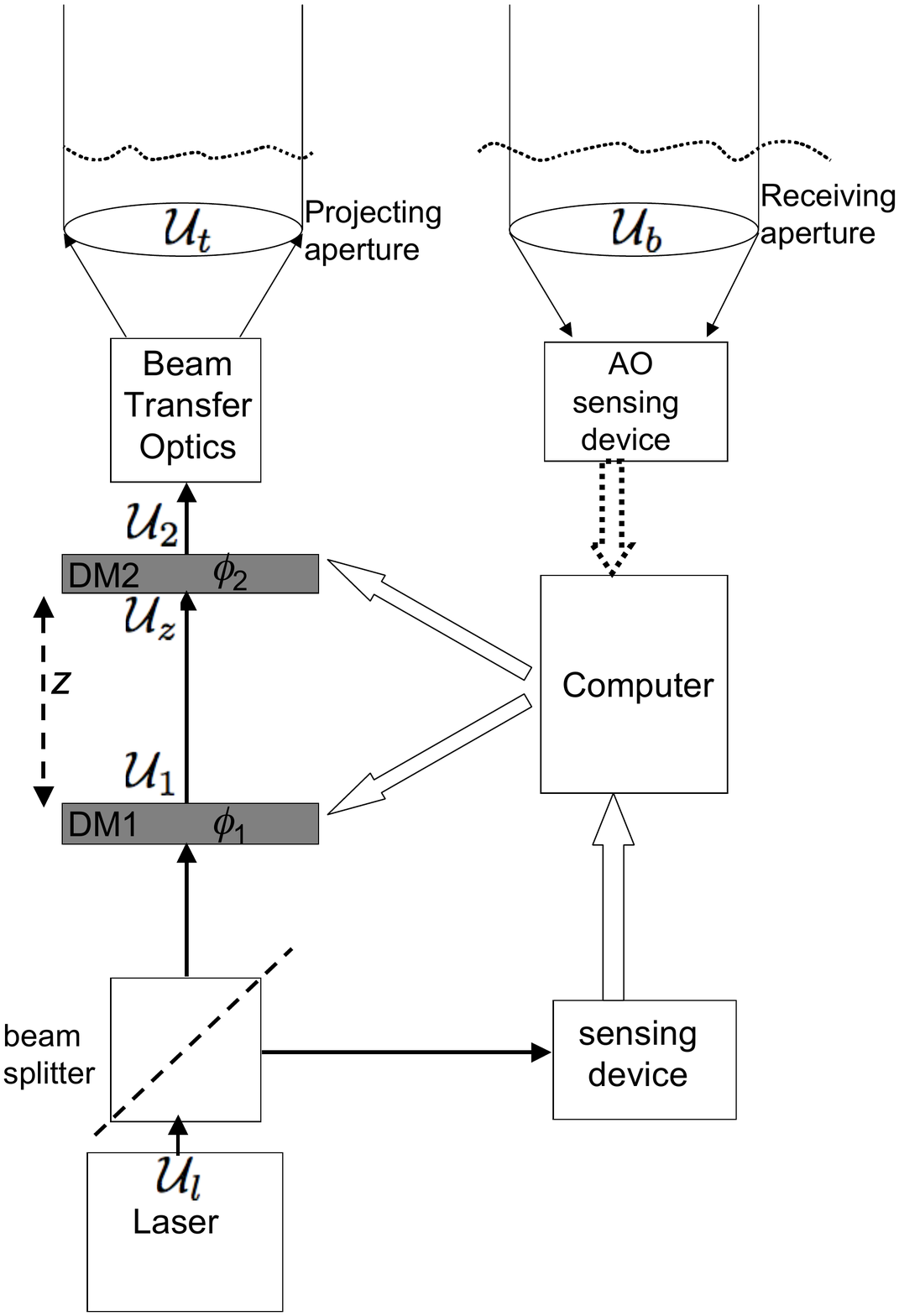}}}
  \caption{\label{fig:2DM-schemes} Schematics of 2-DM correcting
    systems. {\bf (a):} Mono-static system with DM2 in the far-field of
    DM1 \cite{RoggemannLee1998a}. The FTM box represents a Fourier
    transforming mirror. {\bf (b):} 2-DM correcting system in the
    near-field proposed in this paper. DM1 is located in the near
    field of DM2, at a distance $z$. The configuration is named
    \textit{bistatic} for having different receiving and projecting
    apertures.}
\end{figure}

In the fields of defense industry and free space communications, the
beam propagation through the atmosphere leads to strong amplitude
distortions degrading the system performance. During the last two
decades, a significant amount of research was generated for these
applications to develop adaptive optics systems correcting for both
amplitude and phase distortions of a beam
\cite{RoggemannLee1998a,
  BarchersFried2002a,VorontsovKolosov2005a,ZhaoLyke2008a,KizitoRoggemann2004a}
.

One of the first approach to tackle phase and amplitude correction
simultaneously with two DMs was proposed by Kanev
\textit{et al.} \cite{KanevLukin1991a}. The first DM corrects for the
amplitude and the second one corrects for the phase. The concept has
later been revised by Roggemann \textit{et al.}
\cite{RoggemannLee1998a,RoggemannDeng1999a,RoggemannKoivunen2000a} in
a configuration where the second mirror is in the far-field of the
first one. The scheme of such correction system is presented in
Fig.~\ref{fig:2DM-schemes}(a), and it is referred to as
\textit{mono-static} because the receiving and projecting apertures are
the same. The phase and amplitude corruption induced by the atmosphere
in the optical path from the target above the aperture is supposed to
be measured with a suitable device. It could be a SH
sensor, from which outputs need to be processed to compute the
received amplitude and the phase of the field in the aperture plane of
the telescope. The measurements are analyzed to deduce the commands to
be sent to the DMs in order to shape the laser beam to
the conjugate of the received field, and project it. First, the laser
collimated beam is supposed to fall upon DM1, where a phase shape
$\phi_1$ is added to the wave. This applied phase aims at modifying
the amplitude shape of the wave in the Fraunhofer region, after the
Fourier transforming mirror, when the wave is incident on DM2. The
second DM, DM2, is optically conjugated to the pupil plane of the beam
projecting aperture, so it only modifies the phase distribution of the
output beam. A novel modification of the latter approach to near-field
correction has been proposed by Barchers \textit{et al.}
\cite{BarchersFried2002a,BarchersEllerbroek2001a,Barchers2001a,Barchers2002c}
showing the potential interest of conjugating the first DM to a finite
range where a turbulent layer could be present, instead of using a
Fourier transforming mirror. This kind of dual-DM systems to
precompensate atmospheric turbulence effects on beacon has been used
during the last two decades in the military and free-space
communication fields.


In the adaptive optics systems for astronomy, only few studies tackled
full optical field conjugation, because phase correction is by far
more important than amplitude correction \cite{Roddier1999a}. The use
of multiple DMs in AO for astronomy are common for phase correction
only and in tomographic configurations, aiming to correct the phase
over a wide field of view like GeMS
\cite{RigautNeichel2014a}. Nevertheless, amplitude distortions or
\textit{scintillation}, being a source of loss of contrast in
astrophysical imaging, Gonsalves suggested \cite{Gonsalves1997a} to
correct for amplitude distortions thanks to a second DM. Again for
higher contrast purposes, phase-induced amplitude apodization
\cite{Guyon2003a} is an active are of research using aspheric mirrors
for exo-planet imaging. And in the same field, multiple DMs have
recently be shown to provide polychromatic correction and allow to
null the intensity in High Strehl images
\cite{PueyoKasdin2007a,PueyoKay2009a}. With respect to the lasers
for astronomical AO, an on-going project studies the potential benefit
of only phase correction before projection to the mesosphere
\cite{GavelAmmons2008a,NortonGavel2014a}. More recently another
concept of laser beam cleanup with 2 DMs has been presented
\cite{LeiWang2012a} but again it targets only phase distortions since
the second DM is conjugated to the first one and is used only to
increase the maximum spatial frequency of the correction and the
stroke of the overall system. To the knowledge of the authors, the
present study of amplitude and phase correction of the laser beam
before projection for astronomical AO is new.

The 2DM-system proposed here for astronomical purposes uses a second
DM conjugated to a finite range, $z$ (see
Fig.~\ref{fig:2DM-schemes}(b)). DM2 is in the near-field of the DM1
(\textit{i.e.} in the Fresnel region), at a separation distance $z$ of
the order of 1 meter. DM2 remains conjugated to the pupil plane of the
beam projecting aperture as in Fig.~\ref{fig:2DM-schemes}(a)). The
near-field configuration should allow matching the space constraints of
the laser systems on the telescope, although maintaining the ability
to correct for amplitude and phase of the field.

The primary objective of the 2-DM concept is to provide correction of
quasi-static amplitude and phase aberrations affecting the beam before
its projection. The distortions of GeMS laser such as the ones
presented in Fig.~\ref{fig:irrmap} have been observed to only evolve
over a time scale of hours or days \cite{FesquetAraujo2013a}. The
developed solution focuses only on these quasi-static distortions but
the choice of deformable mirrors would allow in a further extension of
the study to tackle optimizations at shorter time scales. For
instance, it may be used to finely tune the focus of the output beam
when variations of the sodium layer height occurs. The DMs could also
precompensate for the dynamical aberrations induced by the atmospheric
turbulence in the propagation path of the laser above the launch
telescope. This last extension is far from being straightforward
however since the turbulence of interest is not directly measured. In
astronomical laser-based systems, the configuration would usually be
\textit{bistatic} (\textit{cf.}  Fig.~\ref{fig:2DM-schemes}(b)),
meaning that the laser launching telescope (left) is not the telescope
measuring the atmospheric turbulence (right). For on-axis launching of
the laser, the turbulence above the secondary mirror, \textit{i.e.}
behind the central obscuration of the laser-based AO system of the
telescope, might be extrapolated for low-order modes corrected by the
ground-conjugated DM of the AO.  The atmospheric wavefront distortion
would thus be known thanks to the adaptive optics of the telescope
(represented by \textit{AO sensing device} in
Fig.~\ref{fig:2DM-schemes}(b)), for which the receiving aperture is
the primary mirror and does not coincide with the launching
telescope. Demonstration of such capability is yet unknown to the
authors, so such dynamical extension of the correction is left for
further research. In Fig.~\ref{fig:2DM-schemes}(b), the output of the
AO sensing device is connected with a dashed arrow to the processing
computer illustrating the fact that this option is not considered
further in this paper.

The correction of quasi-static amplitude and phase distortions is
based on a given desired beam shape at the projection aperture,
$U_t$. The phase patterns to be applied to the DMs are
determined by a phase retrieval method \cite{Fienup1982a} like in most
of the amplitude-phase correction systems mentioned before. The
principle of the method is to iteratively estimate in two planes the
phase shapes which match the amplitude and phase constraints in other
planes (\textit{e.g.} image plane, conjugated plane), as far as they
are linked by a propagation relation. The first phase retrieval
methods were developed by Gerchberg and Saxton
\cite{GerchbergSaxton1972a}, Gonsalves \cite{Gonsalves1976a} and
Fienup \cite{Fienup1978a}. Fienup showed that these methods have the
same local convergence properties around a fix point as the steepest
descent and conjugated gradient algorithms \cite{Fienup1982a}. The
intensive research on optical field conjugation for laser defense
applications led to eventually demonstrate \cite{BarchersFried2002a}
that algorithms for optical field conjugation, usually referred to as
\textit{Gerchberg-Saxton algorithms}, are applications of what is
known as the method of sequential projections onto constraints sets
\cite{StarkYang1998a}. The algorithm presented in Sect.~\ref{sec:Eqs}
is a version of Gerchberg-Saxton method adapted to account for the
near-field separation of the DMs and to unwrap the phases to be
applied on DMs.

Last, to bring the 2DM correction concept to a practical
implementation the method needs to provide phase estimates that can be
efficiently reproduced by the mirrors. Non-segmented mirrors do not
perform well when phase cuts are present
\cite{VenemaSchmidt2008a}. Phase cuts can occur because the phase
estimate is wrapped and because of branch points \cite{Fried1998a}. As
shown later with simulations in Sect.~\ref{sec:simu-field-conj}, the
estimated phases in our context are very likely to be wrapped and to
generate branch points where the field amplitude vanishes in the
aperture. To avoid phase cuts, we developed a regularized iterative
unwrapper (see Sect.~\ref{sec:weight-ls-unwrapper}) to be applied to
the phase estimates after every iteration of the phase-retrieval
algorithm. The use of least squares and weighted least squares
unwrappers have been reported in the literature related to beam
shaping with DMs \cite{Barchers2002c, VenemaSchmidt2008a}. The least
squares approach is not adapted to cases where the field intensity
vanishes. A weighted least squares unwrapper allows to deal with
degeneracies of the phase estimates where the intensity is zero, and
thus avoid phase cuts. However, it is difficult to implement a
weighted least squares unwrapper in practice because the weight based
on the intensity distribution of the computed propagated field changes
at every iteration of the phase retrieval algorithm. So the
pseudo-inverse matrix required to apply the weighted least squares
cannot be precomputed only once. To overcome these difficulties, a
novel iterative unwrapper is detailed in
Sect.~\ref{sec:weight-ls-unwrapper}. Its weighting can be easily
updated on the fly with every new field estimate and no precomputation
of pseudo-inverse matrix is required.

\section{Algorithm for amplitude and phase correction}
\label{sec:Eqs}

For the 2-DM design presented in Fig.~\ref{fig:2DM-schemes}(b), the
following notations are used. The incident field on DM1 is the one
coming out from the laser bench and is noted
\begin{equation}
  \label{eq:Ul}
  U_l(\V{x}) = u_l(\V{x})\, \exp(j\,\phi_l(\V{x}))\,,
\end{equation}
where $\V{x}$ is the coordinate vector in the plane of DM1, and
$u_l(\V{x})$ and $\phi_l(\V{x})$ account for the aberrated amplitude
and phase in this plane respectively. The field leaving DM1 is noted
$U_1$ and is expressed as
\begin{equation}
  \label{eq:U1}
  U_1(\V{x}) = U_l(\V{x}) \, m_1(\V{x}) \, \exp(j\,\phi_1(\V{x}))\,,
\end{equation}
where $\phi_1$ is the phase pattern applied to DM1 and $m_1$
represents the reflection mask for the mirror, in order to account for
the finite extension of the device.

The physics between DM1 and DM2 is modeled with the Fresnel integral
as an approximation of the propagation. This free-space propagation of
the electromagnetic wave $U_1$, at distance $z$, is described by the
linear unitary transformation $T_z$ such that
\cite{BarchersEllerbroek2001a,Schmidt2010a}
\begin{equation}
  \label{eq:Uz}
  U_z(\V{x}')  = {\rm T}_z\left[U_1(\V{x})\right] = \mathcal{F}^{-1} \left[ \mathcal{F}(U_1) \,\, \exp(-j \pi \lambda z |\V{f}_x|^2) \, \exp(j 2 \pi z / \lambda) \right]\,,
\end{equation}
where $\V{f}_x$ is the spatial frequency vector, $\lambda$ is the
laser wavelength and $U_z$ is the incident field on DM2. After DM2
phase correction $\phi_2$, the output field is 
\begin{equation}
  \label{eq:U2}
  U_2(\V{x}') = U_z(\V{x}') \, m_2(\V{x}')\, \exp(j \, \phi_2(\V{x}'))\,,
\end{equation}
where $m_2$ represents the mask of this mirror pupil. DM2
is conjugated to the projecting aperture such that $U_2$ is assumed to
be the transmitted field.

In our context, the optical field desired to be transmitted by the
launching aperture is assumed known, and we refer to it as the desired
output field $U_t$, noted
\begin{equation}
  \label{eq:Ut}
  U_t(\V{x}') = u_t(\V{x}') \, \exp(j\, \phi_t(\V{x}'))\,,
\end{equation}
where $u_t$ is its amplitude and $\phi_t$ its phase. This desired
output field is a constraint in the algorithm. Its choice is discussed
in \cite{BechetGuesalaga2013a} and in the present paper we only
consider it as being known and fixed.

In laser applications where strong scintillation is produced by the
turbulence during the propagation above the launching aperture, the
transmitted field $U_t$ needs to be the conjugate of the field which
would be received in a mono-static configuration like in
Fig.~\ref{fig:2DM-schemes}(a) \cite{BarchersEllerbroek2001a}. In
the astronomical context, the weak turbulence above the telescope
leads to negligible amplitude distortions, such that the sensing
device located after the receiving aperture in
Fig.~\ref{fig:2DM-schemes}(b) is aimed at analyzing the phase
distortions $\phi_b$ only.

Assuming $U_t$ and $U_l$ known, the goal of the 2DM concept is to
determine $\phi_1$ and $\phi_2$ to be applied to DM1 and DM2
respectively in order to obtain the field $U_t$ on the projecting
aperture \cite{RoggemannLee1998a}. For this, a phase retrieval method
is used, iteratively providing estimates of $\phi_2$ and $\phi_1$ to
satisfy the intensity constraints fixed by $U_l$ and $U_t$
respectively. The main steps of the algorithm, similar to the one
described by Roggemann and Lee \cite{RoggemannLee1998a} are described
in Sects.~\ref{sec:iterate-phi2} and \ref{sec:iterate-dm1}. In
practice, the algorithm must also unwrap the phase estimates in order
to guarantee efficient continuous-mirror correction. A specific
iterative unwrapper is developed for this purpose in
Sect.~\ref{sec:weight-ls-unwrapper}.

\subsection{Forward iterative step for DM2}
\label{sec:iterate-phi2}

Given an initial
estimate for $\phi_1$, the incoming field on DM2 is computed
\begin{equation}
  \label{eq:onDM2}
  U_z(\V{x}') = {\rm T}_z \{ u_l(\V{x}) \, m_1(\V{x}) \, \exp[j\, (\phi_l(\V{x}) + \phi_1(\V{x}))] \}\,.
\end{equation}
Since we want the field after DM2 to be equal to the desired output
$U_t$, the required phase to apply on DM2 $\phi_2$ is constrained to
satisfy
\begin{equation}
  \label{eq:phi2-constr}
  \phi_2(\V{x}') = - \arg \left[ U_t^*(\V{x}') \, U_z(\V{x}') \,m_2(\V{x}')\right]\,.
\end{equation}

Once the phase $\phi_2$ is obtained from Eq.~(\ref{eq:phi2-constr}),
we compute an additional parameter $\alpha$, which is
the ratio of the incident energy $E_2$ on DM2 over the energy $E_t$ of
the desired output field $U_t$. This factor $\alpha$ is used later
in Eq.~(\ref{eq:Uz-back}) to rescale the field to be propagated
backward in the iterative step for determination of $\phi_1$. The
value of $\alpha$ is mathematically defined by
\begin{equation}
  \label{eq:alpha}
  \alpha = \frac{E_2}{E_t} = \frac{\int \int \Abs{U_z(\V{x}')\,m_2(\V{x}')}^2 {\rm d}\V{x}'} {\int \int \Abs{U_t(\V{x}')\,m_2(\V{x}')}^2 {\rm d}\V{x}'}\,.
\end{equation}

\subsection{Backward iterative step for DM1}
\label{sec:iterate-dm1}

This second step estimates the phase $\phi_1$ to be applied to DM1
\cite{RoggemannLee1998a}. It is obtained by back propagation of the
rescaled desired output field. The incident field on DM2 is thus
\begin{equation}
  \label{eq:Uz-back}
  U_z(\V{x}') = \sqrt{\alpha}\, \, U_t(\V{x}') \, m_2(\V{x}') \, \exp(-j \phi_2(\V{x}'))\,,
\end{equation}
and the input field on DM1 can be written as
\begin{equation}
  \label{eq:U-in}
  U_1(\V{x}) = {\rm T}_{-z}\left[ U_z(\V{x}') \right] = {\rm T}_{z}^*\left[ U_z^*(\V{x}')\right]\,.
\end{equation}
The last equality is obtained by the property of the adjoint
operator ${\rm T}_z^*$ of ${\rm T}_z$. Indeed, for any given complex field $U$,
${\rm T}_z^*(U)={\rm T}_{-z}(U^*)$. 

The estimation of phase $\phi_1$ is obtained applying the
constraint that the phase and amplitude after backpropagation beyond DM1
matches the input laser field phase $\phi_l$ and amplitude
$u_l$. Thus,
\begin{equation}
  \label{eq:phi1-constr}
  \phi_1(\V{x}) = \arg \left[U_l^*(\V{x}) \,m_1(\V{x})\,  U_1(\V{x}) \right]\,.
\end{equation}

Equations~(\ref{eq:onDM2})-(\ref{eq:alpha}) and
Eqs.~(\ref{eq:Uz-back})-(\ref{eq:phi1-constr}) represent the core of
the 2-DM correction algorithm, being computed alternatively and
iteratively, in order to obtain the best amplitude and phase
outputs.

It is important to observe that the computation of
Eqs.~(\ref{eq:phi2-constr}) and~(\ref{eq:phi1-constr}) leads to
wrapped phase values. In addition, the field amplitude for which the
argument must be obtained can be null or numerically very close to
zero at some locations within the DMs aperture, which could lead to
undesired branch points \cite{Fried1998a} appearing in the phase
estimations of $\phi_1$ and $\phi_2$. In order to obtain phase
functions $\phi_1$ and $\phi_2$ likely to be efficiently reproduced by
the continuous sheet of a DM, the phases must be
unwrapped and branch points must be avoided.

\subsection{Phase unwrapping}
\label{sec:weight-ls-unwrapper}

In order to overcome these difficulties of wrapped phase and branch
points, an iterative and regularized 
unwrapper estimator is presented. 
This unwrapper is modified from the weighted least squares unwrapper
presented in \cite{GuesalagaNeichel2012a}, and is obtained by
iteratively solving the linear system
\begin{equation}
  \label{eq:Unwrapper}
  \left( \M{G}\T \M{W}_g \M{G} + \mu \,\M{C}\T\, \M{C}\right)\,\hat{\phi} = \M{G}\T \M{W}_g \mathcal{PV}[\M{G} \phi]\,,
\end{equation}
where $\M{G}$ is the discrete representation of the gradient operator
as described by Fried \cite{Fried1998a}. The $\mathcal{PV}$ symbol
stands for the principal-value operator which output is an angle that
falls in the range $\pm\pi$ (modulus $2\pi$), $\M{W}_g$ is a diagonal
array of weights, $\M{C}$ is a discrete representation of the
curvature operator and $\mu$ is a scalar weighting the curvature
regularization term $\M{C}\T \M{C}$.

A wrapped phase has the gradient of its unwrapped phase plus or minus
a multiple of $2\pi$. This is why the unwrapper in
Eq.~(\ref{eq:Unwrapper}) involves the gradient operator on both sides
of the equation and the principal value operator on the right hand
side where the wrapped phase is. In addition, $\phi_2$ and $\phi_1$
are computed from Eqs.~(\ref{eq:phi2-constr}) and
(\ref{eq:phi1-constr}) only where the field amplitude $u$ in the
argument is numerically non zero. This leads in both cases to wrapped
phase maps $\phi$, and the wrapped phase samples are arbitrarily set
to zero where the amplitude is also zero, \textit{i.e.} where
degeneracy occurs.

The matrix $\M{W}_g$ is used as a diagonal weighting related to the
level of confidence of the gradient value of the wrapped phase,
\textit{i.e.}  accounting for the fact that the computed phase and gradients
values do not represent the true values where the field
amplitude reaches zero. The weighting is defined differently
for gradient in $x$ and gradient in $y$, such that the component in
$(i,j)$ associated to gradient along $x$ at location $(x_i, y_j)$, can
be mathematically described as
\begin{equation}
  \label{eq:Wg}
  \M{W}_{gx}(i,j) = \log(u(x_{i+1},y_{j}) \, u(x_{i},y_{j}) + \epsilon) - \log(\epsilon)
\end{equation}
with $\epsilon$ a small numerical value, \textit{e.g.}  $10^{-23}$,
representing the threshold of the product $u(x_{i+1},y_{j}) \,
u(x_{i},y_{j})$ of two neighbour amplitude samples above which the
associated gradient is considered with a strictly positive
weight. This $\epsilon$ is chosen as a negligible value compared to
the average field amplitude over the aperture. This logarithmic scale
has the benefit to spread the weighting values of $\M{W}_g$ mainly in
the range 1 to 10 where non negligible amplitude values have been
obtained. The weight become significantly lower as soon as one field
amplitude involved in the gradient computation is zero, revealing the
proximity of a non-illuminated area. The weight is exactly zero where
the gradient computation involves two field samples of null
amplitude. With such approach, the weighting matrix $\M{W}_g$ is
updated on the fly, everytime the unwrapper is called with a new
wrapped phase and amplitude field. This update is easily handled since
there is no requirement to precompute a matrix inverse to apply the
iterative unwrapper.

Another interesting feature of the unwrapper is the regularizing term
$\M{C}\T \, \M{C}$ (in Eq.~(\ref{eq:Unwrapper})), which favors
solutions of $\hat{\phi}$ with low curvature. The relative importance
of this penalty term can be tuned thanks to the scalar $\mu$ in
Eq.~(\ref{eq:Unwrapper}). In practice, we have found that $\mu$ of the
order of $10^{-3}$ is apropriate to influence the solution only in
areas of negligible amplitude.

\subsection{Convergence criteria}
\label{sec:Conv-crit}

The convergence of the beam shaping iterative method is assessed
thanks to two complementary criteria. The first one $\epsilon^2_{\rm
  ampl}$ measures the relative distance between the currently achieved
output amplitude shape and the desired one
\cite{RoggemannLee1998a}. After $i$ iterations of the method,
\begin{equation}
  \label{eq:crit-dist}
  \epsilon^2_{\rm ampl}(i) = \frac{\int \int \left[|U_t(\V{x}')| - |U_2^{(i)}(\V{x}')| \right]^2 \, {\rm d}\V{x}'} {\int \int \left[|U_t(\V{x}')| - |U_2^{(0)}(\V{x}')| \right]^2 \, {\rm d}\V{x}'}\,
\end{equation}
where the notation $U_2^{(i)}$ stands for the achieved output field
after DM2, when the $i$-th estimations of $\phi_1$ and $\phi_2$ are
applied to DM1 and DM2 respectively. The scalar $\epsilon^2_{\rm
  ampl}$ thus represents a relative amplitude error with respect to
the initial amplitude error.

The second criterion is denoted as $S$ and it is sensitive to the
phase matching between the reached output field and the desired output
one. It is written
\begin{equation}
  \label{eq:crit-S}
  S(i) = \frac{\Abs{\int \int U_t^*(\V{x}') \, U_2^{(i)}(\V{x}') {\rm d}\V{x}'}^2} {\Abs{\int \int U_t(\V{x}')\,U_t^*(\V{x}') {\rm d}\V{x}'}\, \Abs{\int \int U_2^{(i)}(\V{x}')\,U_2^{(i)*}(\V{x}') {\rm d}\V{x}'}}\,,
\end{equation}
where index $(i)$ again refers to the $i-$th iteration of the
procedure. This criterion is inspired from the far-field intensity
criterion used in the literature
\cite{BarchersEllerbroek2001a,BecknerOesch2007a}.

These two criteria are used in the simulations of
Sect.~\ref{sec:simu-field-conj} to enhance the convergence of the
algorithm. They both evolve in the range $[0,1]$. $\epsilon^2_{\rm
  ampl}$ starts at 1 by definition and decreases toward 0 when the
correction improves. On the contrary, $S$ is expected to increase
with iterations toward unity.

\section{Simulations}
\label{sec:simu-field-conj}

In this section, we present numerical simulations of the optical field
conjugation method proposed in the previous section. In a first step,
we check that the optimization of the amplitude and phase outputs is
effective, following the algorithm presented in
Sect.~\ref{sec:Eqs}. Next, the influence on performance of system
parameters is studied: the separation distance $z$ between the
mirrors; the number of iterations; and the grid sampling step of the
propagated fields and phase estimates.

\subsection{Amplitude and phase correction}
\label{sec:ampl-phase-corr}

As a first step, we consider an arbitrary amplitude distortion of the
incoming laser \cite{GuesalagaNeichel2012a}. No phase aberrations are
considered, but a particular output phase shape, a defocus, is
specified. The laser wavelength is 589~nm and the distance between the
DMs is chosen to be $z=1.8~$m.

\begin{figure}[htb!]
  \centering
  \mbox{\subfigure[]{\includegraphics[width=0.35\linewidth]{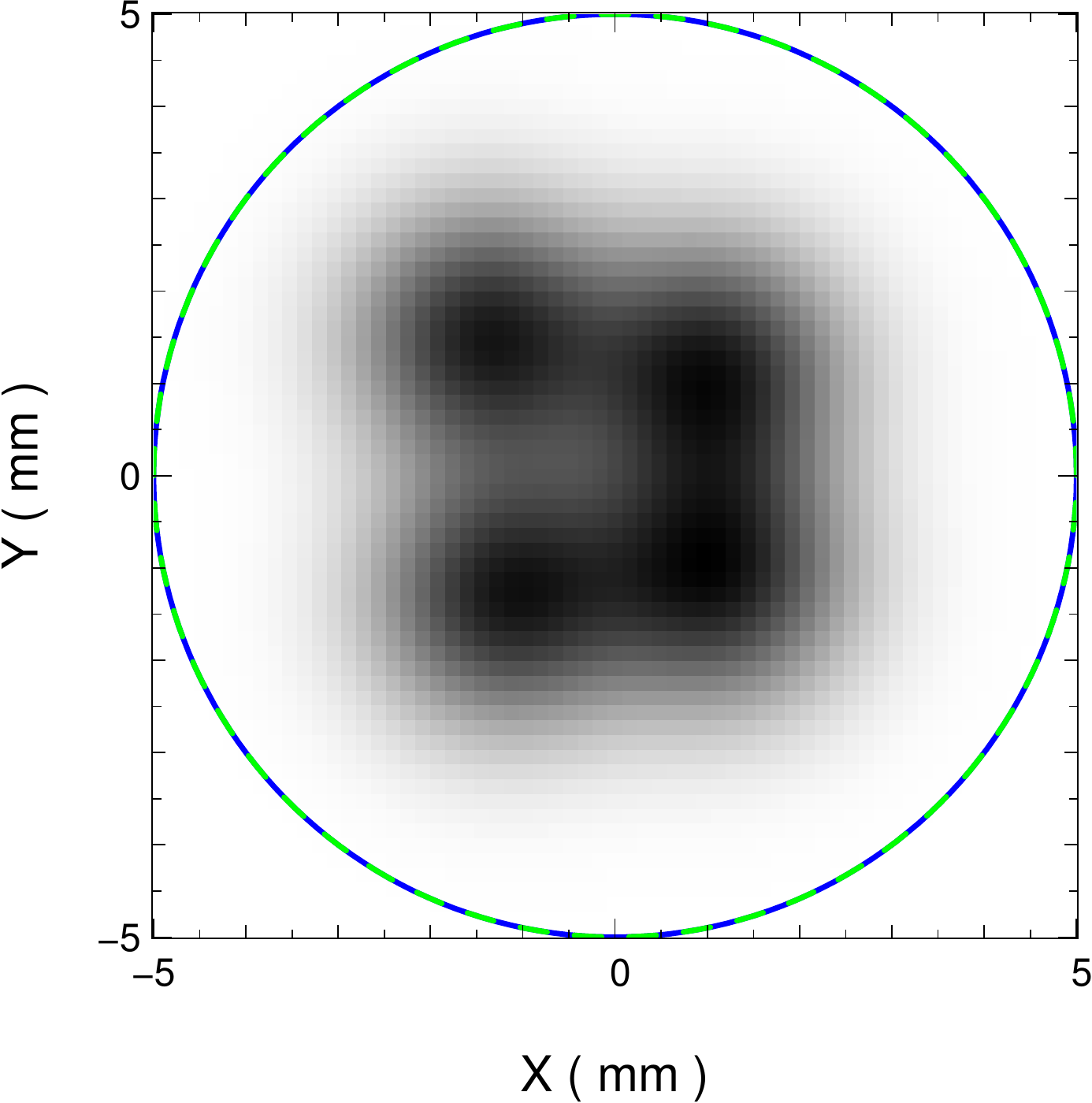}} \hspace{0.5cm}
  \subfigure[]{\includegraphics[width=0.35\linewidth]{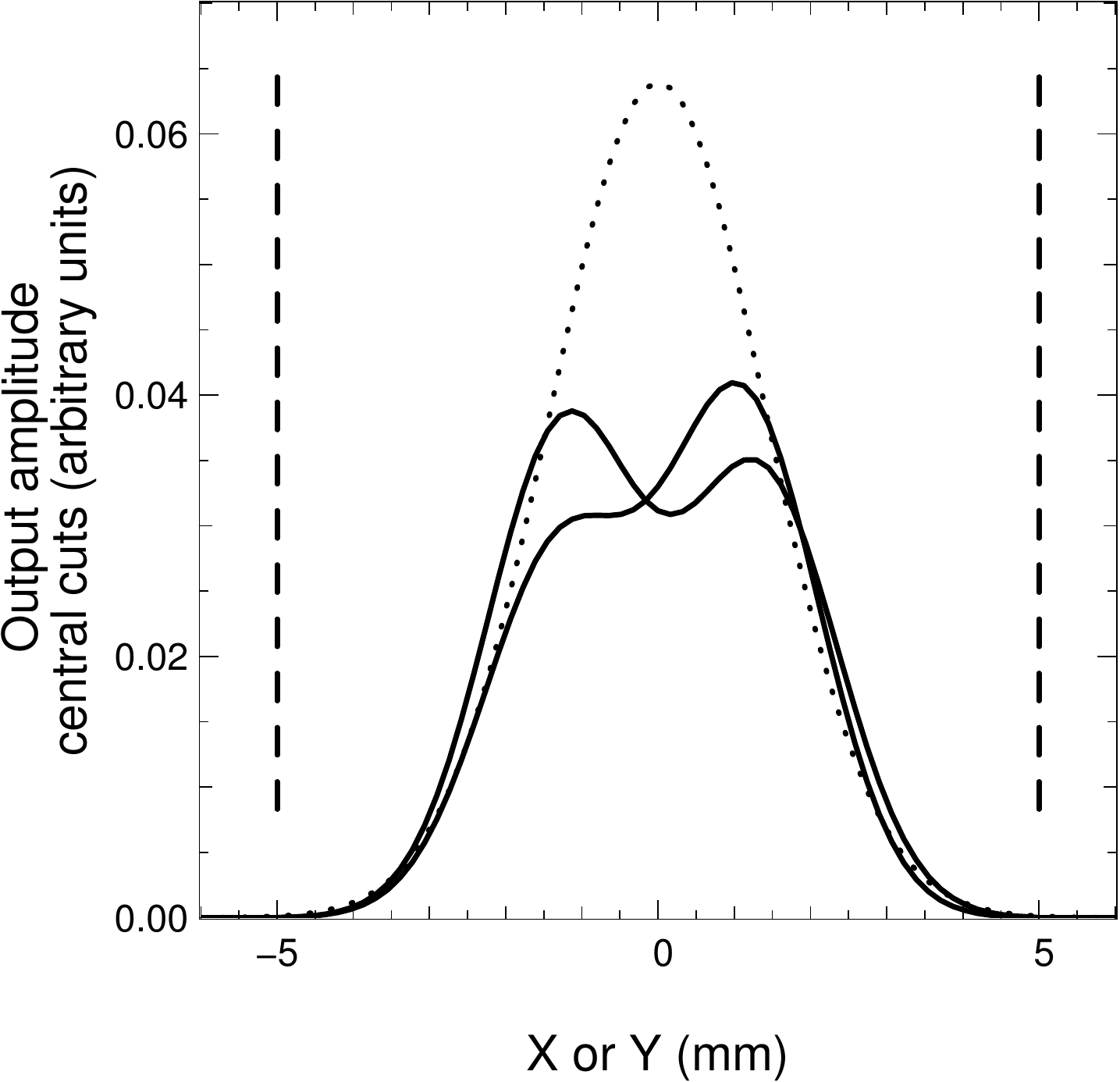}}} 
  \mbox{\subfigure[]{\includegraphics[width=0.35\linewidth]{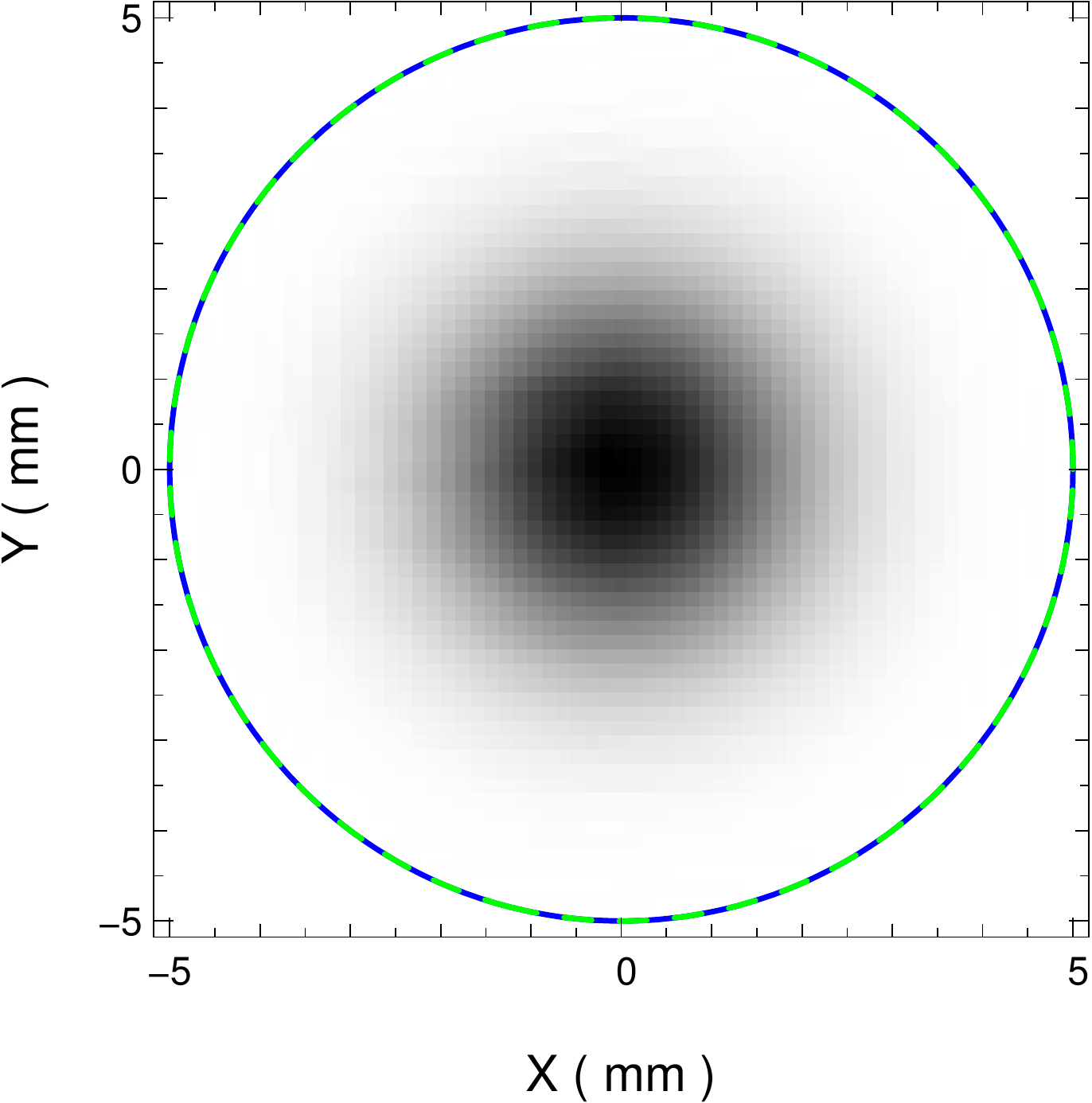}} \hspace{0.5cm}
  \subfigure[]{\includegraphics[width=0.35\linewidth]{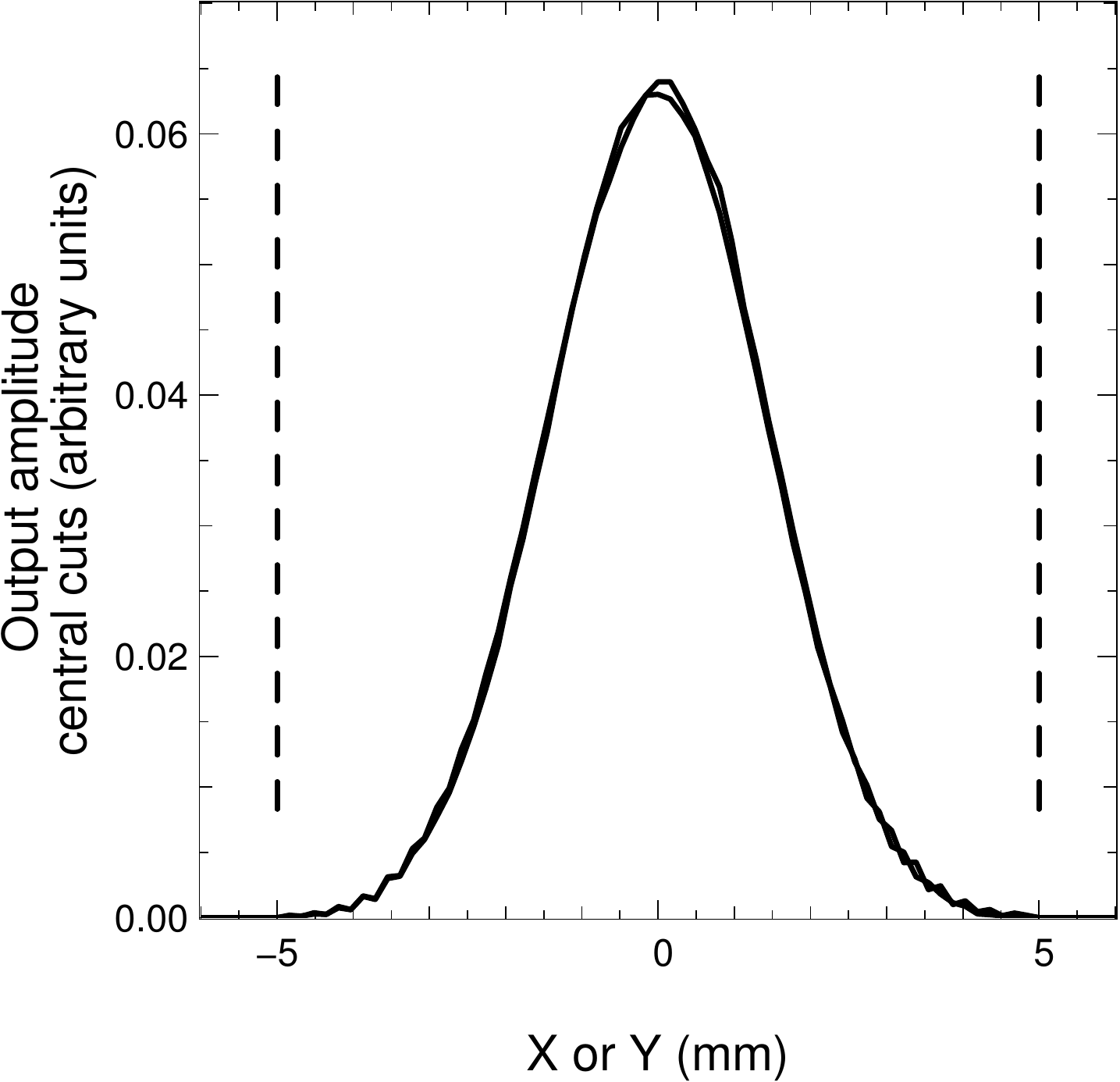}}}
\caption{Amplitudes of the output field at DM2, without correction
  ((a) and (b)) and with correction ((c) and (d)). Graphics (a) and
  (c) represent the 2D amplitude over the 1~cm aperture of
  DM2. Corresponding central cuts along $x$ and $y$ axis are presented
  as solid curves on the associated graphics on the right, (b) without
  correction and (d) with correction. The dotted curve in (b) stands
  for the Gaussian desired output amplitude. In (b) and (d), the
  vertical dashed lines stand for the diameter of DM2.}
\label{fig:output-amplitudes}
\end{figure}

Without any beam shaping correction, \textit{i.e.} $\phi_1 = \phi_2 = 0$, the
input distortions lead to output amplitude very similar to the input
one, slightly modified by the propagation in the near-field
($z=1.8~$m). This output amplitude is represented in
Fig.~\ref{fig:output-amplitudes}(a), over the 1~cm diameter of the
aperture of DM2. Cuts along $x$ and $y$ central axes are represented with
solid curves in Fig.~\ref{fig:output-amplitudes}(b). It can be
observed that the amplitude of the beam is far from being Gaussian.

\begin{figure}[htb!]
  \centering
  \mbox{\subfigure[]{\includegraphics[width=0.35\linewidth]{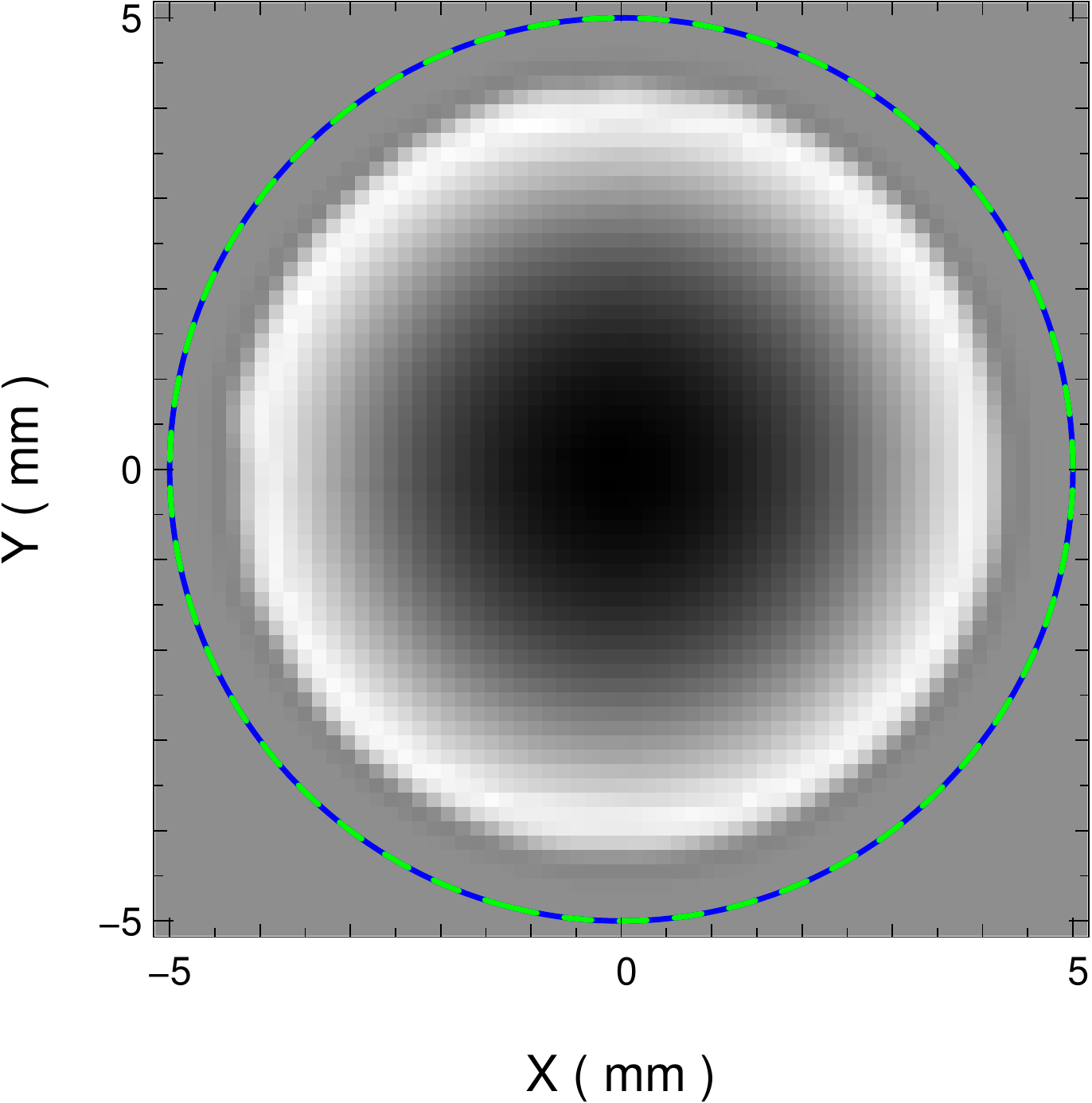}} \hspace{0.5cm}
  \subfigure[]{\includegraphics[width=0.35\linewidth]{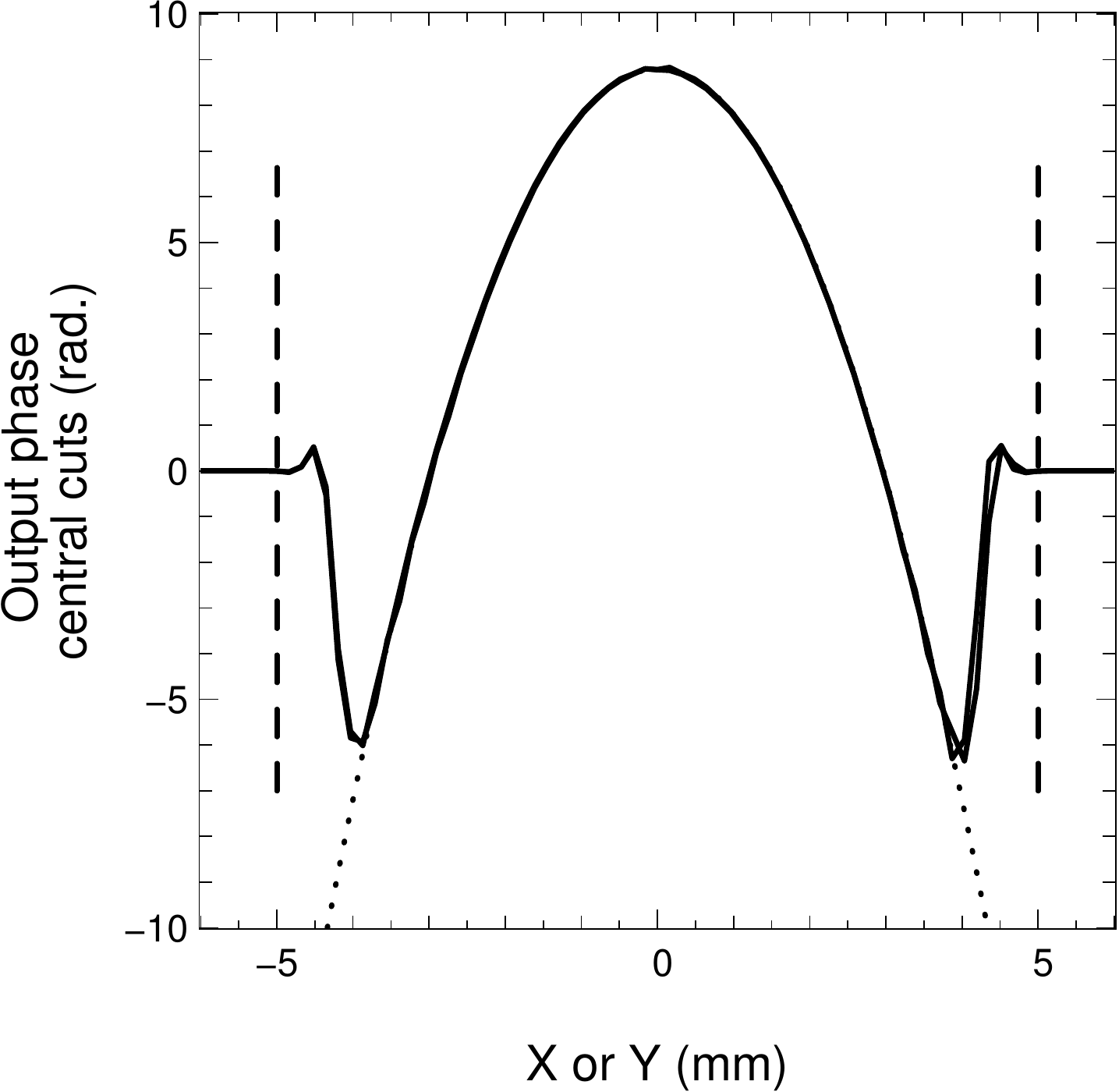}}}
\caption{Estimated phase of the output field after DM2 with
  correction. Graphics (a) represents the 2D phase as estimated over
  the 1~cm aperture of DM2. The central cuts along $x$ and $y$ axis are
  shown with solid curves in (b). The dotted curve in (b) stands for
  the cuts of the desired output phase (defocus), which coincides with
  the achieved output in areas of non-negligible amplitude (see (d) in
  Fig.~\ref{fig:output-amplitudes}). The vertical dashed lines in (b)
  delimit diameter of DM2.}
\label{fig:output-phase}
\end{figure}

The desired output amplitude is set to a Gaussian shape,
represented by dotted slices along $x$ and $y$ central axes in
Fig.~\ref{fig:output-amplitudes}(b). The desired output phase has been
arbitrarily chosen to be a defocus (see dotted curve in
Fig.~\ref{fig:output-phase}(b)).

\begin{figure}[htb!]
  \centering
  \includegraphics[width=0.4\linewidth]{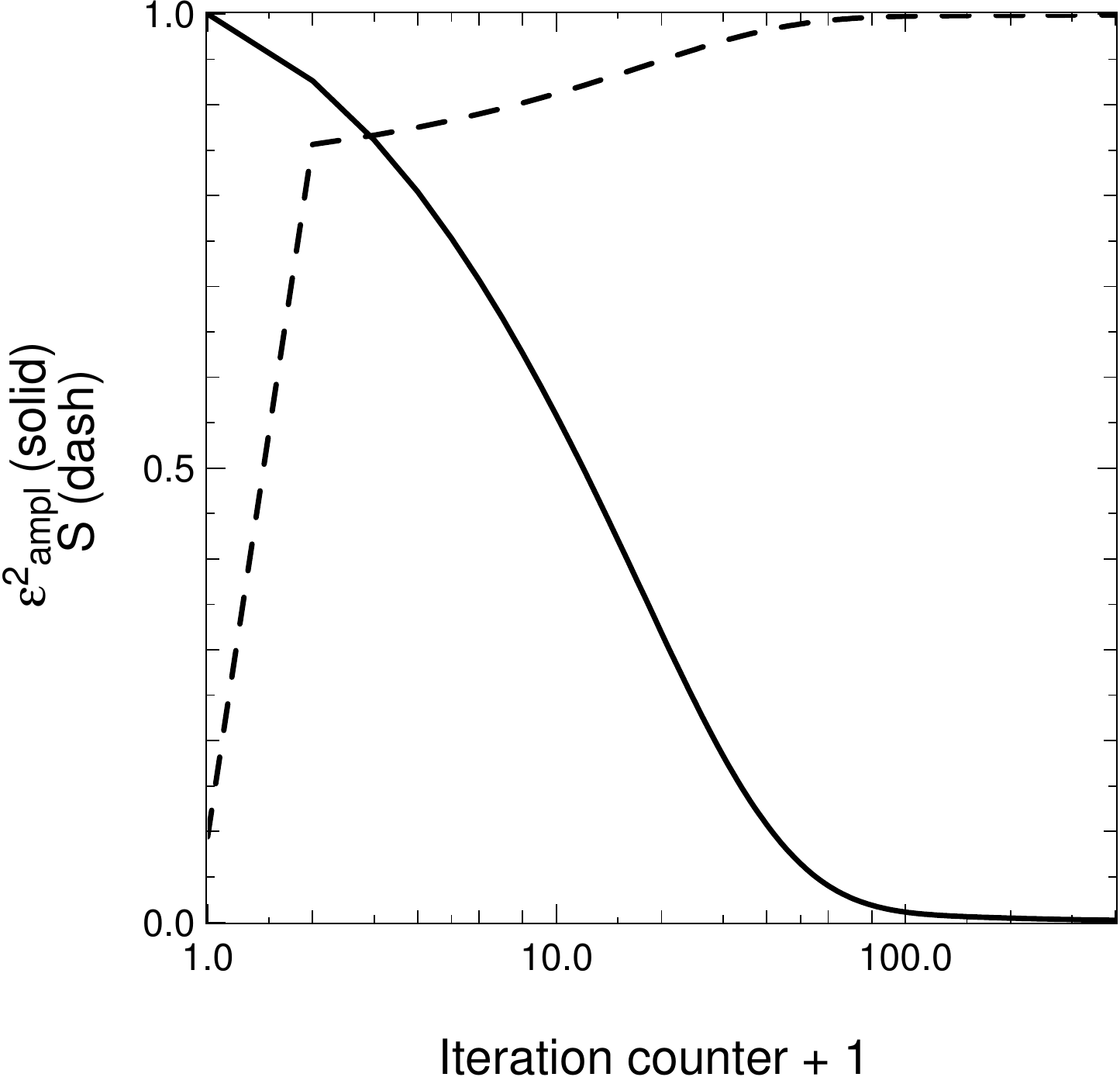}
  \caption{Convergence of the criteria $\epsilon^2_{\rm ampl}$ (solid)
    and $S$ (dashed) defined in Eqs.~(\ref{eq:crit-dist}) and
    (\ref{eq:crit-S}). The iteration counter in abscissae is presented
    in logarithmic scale from 1 to 400.}
\label{fig:test4-conv}
\end{figure}

After 400 iterations of the phase retrieval method presented in
Sect.~\ref{sec:Eqs}, the output amplitude and phase after DM2 almost
perfectly match the desired ones. The two-dimensional representation
of the output amplitude is shown in
Fig.~\ref{fig:output-amplitudes}(c), and the one of the output phase
in Fig.~\ref{fig:output-phase}(a). The central slices along $x$ and $y$
are plotted with solid curves, in Fig.~\ref{fig:output-amplitudes}(d)
for the amplitude and in Fig.~\ref{fig:output-phase}(b) for the
phase. The convergence of the algorithm along the 400 iterations is
illustrated in Fig.~\ref{fig:test4-conv}. Both corrections of
amplitude and phase work perfectly.

It is worth noting that output phase estimated in
Fig.~\ref{fig:output-phase} differs from the desired phase shape near
the aperture edges. This is due to the fact that the dotted curve
represents a given phase function, the desired output. On the
contrary, the solid curve represents an estimate of the resulting
phase, computed from the simulation of the field
propagation. Therefore, in the same way as explained in
Sect.~\ref{sec:weight-ls-unwrapper} for $\phi_1$ and $\phi_2$
computations, the estimation of the resulting phase is initially
wrapped and only at non zero amplitude locations. The field amplitude
(shown in Fig.~\ref{fig:output-amplitudes}(c)) for radii greater than
4~mm is so small that the phase cannot be estimated with accuracy at
these locations. The solid curve in Fig.~\ref{fig:output-phase} is
thus the result of the iterative unwrapper applied to the output phase
estimate. In the outer ring area, the regularization of the iterative
phase unwrapper detailed in Sect.~\ref{sec:weight-ls-unwrapper} mainly
drives the estimation of the phase. However, whatever the phase
estimate in this area, it has no effect on the beam shaping
performance because the field amplitude is almost zero. Note also that
the criterion $S$ in Fig.~\ref{fig:test4-conv} is not notably affected
by the phase discrepancy in this area because of the amplitude
weighting in Eq.~(\ref{eq:crit-S}).

From Fig.~\ref{fig:test4-conv}, it can be concluded that the beam
shaping has worked to perfection after 150 iterations. The correction
of the output phase of the beam is almost instantaneous, as
demonstrated by the value of $S$ already jumping to 85\% after a
single iteration. Amplitude correction drives the global
convergence. The influence of the DMs separation distance on the
convergence speed is further studied in Sect.~\ref{sec:propag-z}.

\begin{figure}[htb!]
  \centering
  \includegraphics[width=0.4\linewidth]{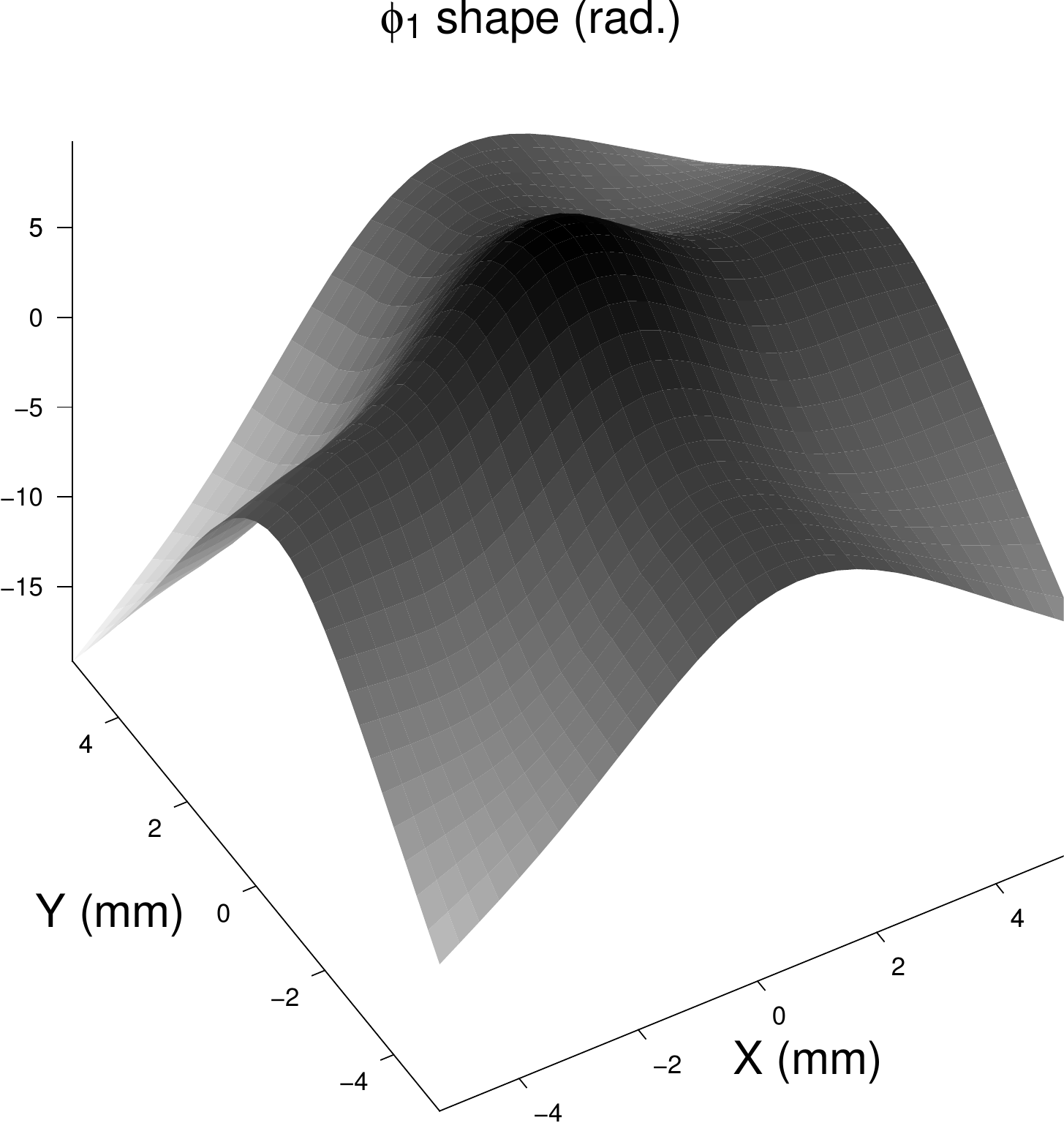}
  \hspace{0.5cm}
  \includegraphics[width=0.4\linewidth]{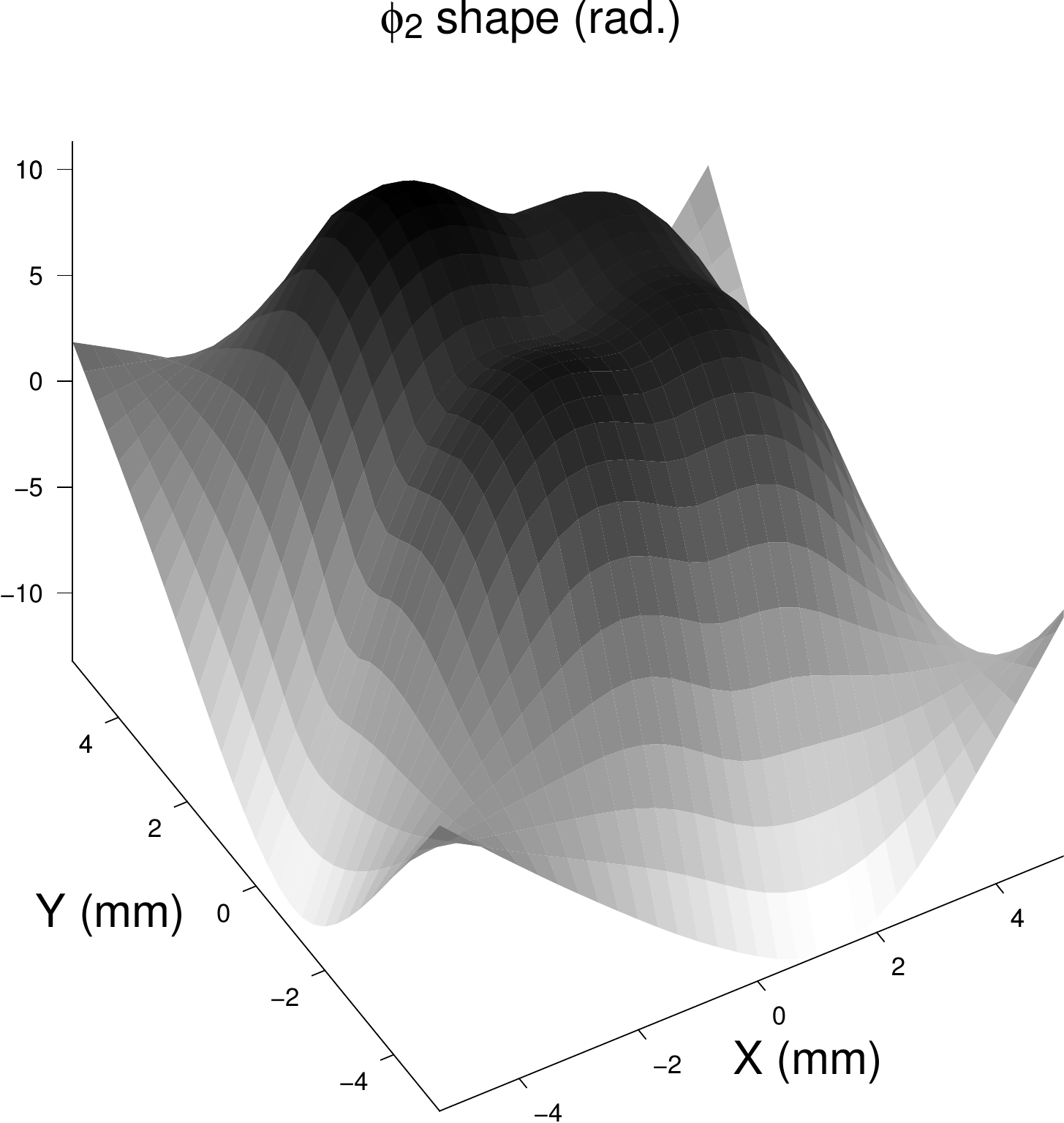}
  \caption{Phase corrections $\phi_1$ ({\bf left}) and $\phi_2$ ({\bf
      right}) to be applied to DM1 and DM2 respectively, after 400
    iterations.}
\label{fig:test4-phi1-phi2}
\end{figure}

In this first simulation, the grid sampling size of $\phi_1$ and
$\phi_2$ is identical to the resolution $\delta$ of the introduced
aberrations $U_l$ and of the desired output field $U_t$, with
$\delta\simeq 161~\mu$m. The desired output amplitude is a perfect
Gaussian with FWHM of 1.7~mm, and the DMs pupils are
considered to be 1~cm in diameter. The phase distortions $\phi_1$ and
$\phi_2$ which should be applied on DM1 and DM2 in order to obtain
such correction are presented in Fig.~\ref{fig:test4-phi1-phi2}.

These smooth shapes are otbained thanks to the regularized unwrapper
described in Sect.~\ref{sec:weight-ls-unwrapper}, applied to the
wrapped phases computed from iterative procedures of
Sects.~\ref{sec:iterate-phi2} and \ref{sec:iterate-dm1}. 

\begin{figure}[htb!]
  \centering
  \includegraphics[width=0.4\linewidth]{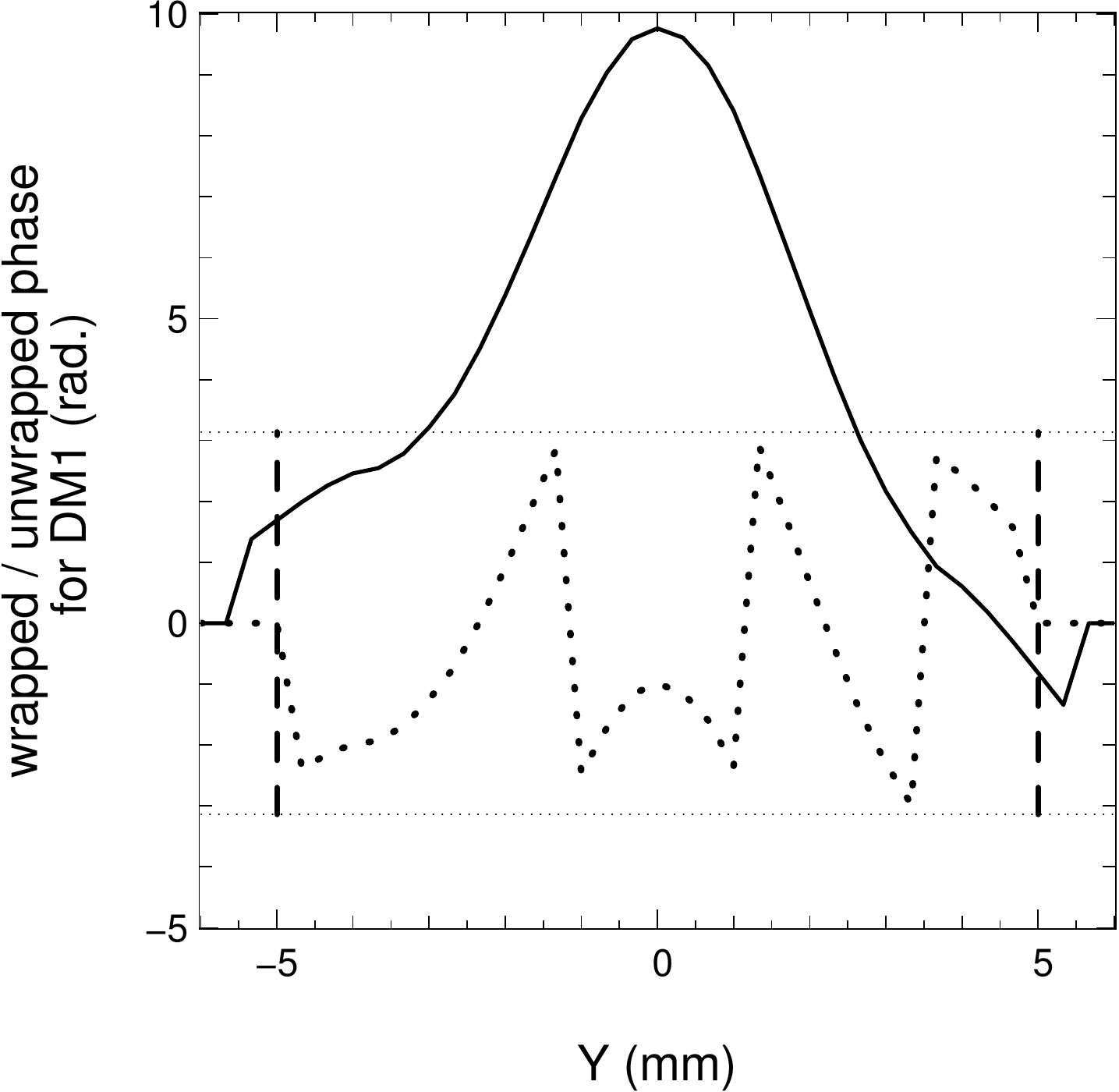}
  \caption{Central cut along $y$ for the final wrapped (dotted curve) and
    unwrapped (solid curve) $\phi_1$ estimate shown in
    Fig.~\ref{fig:test4-phi1-phi2}. The horizontal dots represent the
    $]-\pi;\pi]$ interval of the wrapped phase estimation. The
    vertical dashed lines stand for the diameter of DM2. }
\label{fig:test4-phi1-wrapped}
\end{figure}

The action of the phase unwrapper is illustrated in
Fig.~\ref{fig:test4-phi1-wrapped}, where the central cuts along $y$
for $\phi_1$ is shown in dotted line for the wrapped solution (between
$-\pi$ and $\pi$) and in solid line once unwrapped. The limit of the
mirrors pupil diameter are represented by the vertical dashed
lines. Again it can clearly be noticed that in the center of the slice
(where the amplitude of the field is larger), the unwrapper accurately
unwraps the phase estimates. On the pupil edges however, where the
amplitude vanishes, the regularizing term of the unwrapper becomes
significant and the unwrapped estimate is extrapolated, penalizing
local curvature of the phase.

\subsection{Influence of DM separation}
\label{sec:propag-z}

In this section, the influence of the separation distance between the
2 DMs is investigated. The idea of the near-field position of DM2
comes from possible space constraints on the telescope. The simulations
presented in this section demonstrate that such near-field propagation
is able to produce satisfactory amplitude corrections for the laser
beam.

\begin{figure}[htb!]
  \centering
  \includegraphics[width=0.4\linewidth]{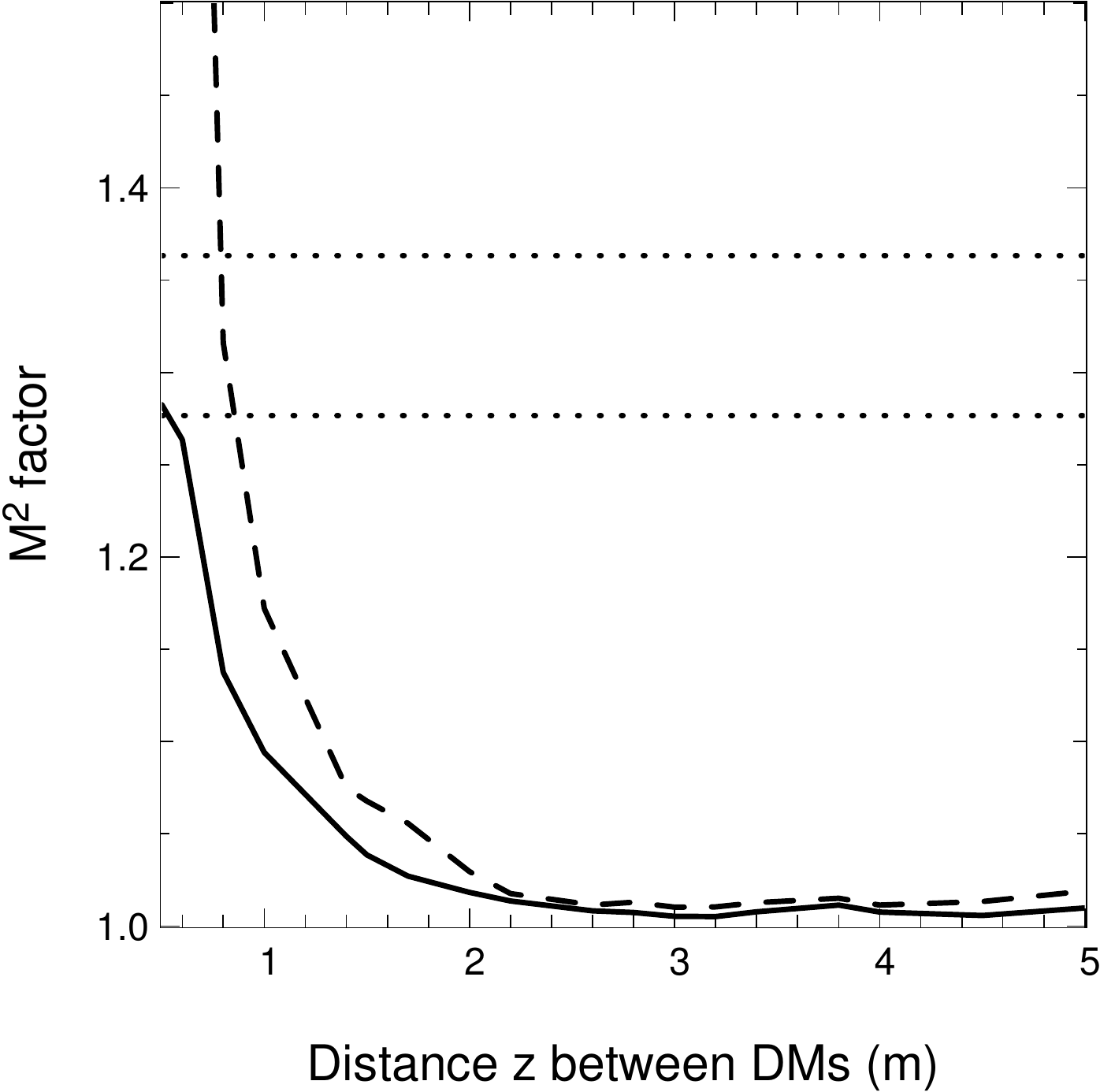}
  \caption{Impact of the separation distance between the DMs on the
    $M^2$ factor of the output laser beam after 400 iterations. Solid
    line: $M^2_x$. Dashed line : $M^2_y$. $M^2_x$ and $M^2_y$ of the
    input laser beam are represented by the horizontal dotted lines
    ($M^2_x$: top; $M^2_y$ : bottom).}
\label{fig:test3-varying-z}
\end{figure}

The simulated system is identical to the one presented in
Sect.~\ref{sec:ampl-phase-corr}, but a flat output phase is sought, in
order to focus the attention on the amplitude correction
abilities. The final correction performance is evaluated in terms of
the $M^2$ factor of the output beam. This criterion is applicable in
this case because the desired output amplitude is a Gaussian
shape. The results of the correction performance after 400 iterations
are presented in Fig.~\ref{fig:test3-varying-z}.

The output $M^2_x$ and $M^2_y$ are shown in solid and dashed lines as
a function of the separation distance $z$ between the DMs. The dotted
horizontal lines stand for the $M^2_x$ and $M^2_y$ values when no beam
shaping correction system exists. These values are computed assuming
that the output beam is $U_l$. When the beam shaping system is
considered, the $M^2$ factor is computed after the DM2, which means
that even if no DM correction is applied, the output beam is slightly
affected in the near-field. Note that for $z\leq 1~$m, $M^2_y$ factor
achieved after 400 iterations is worse than when no correction is
applied. This highlights the difficult amplitude correction at short
DMs separation. For distances greater than $2~$m, it can be observed
that the beam correction is very efficient, with resulting $M^2$
factor values as low as 1.02. Such values mean very good quality
Gaussian beams.

\begin{figure}[htb!]
  \centering
  \includegraphics[width=0.4\linewidth]{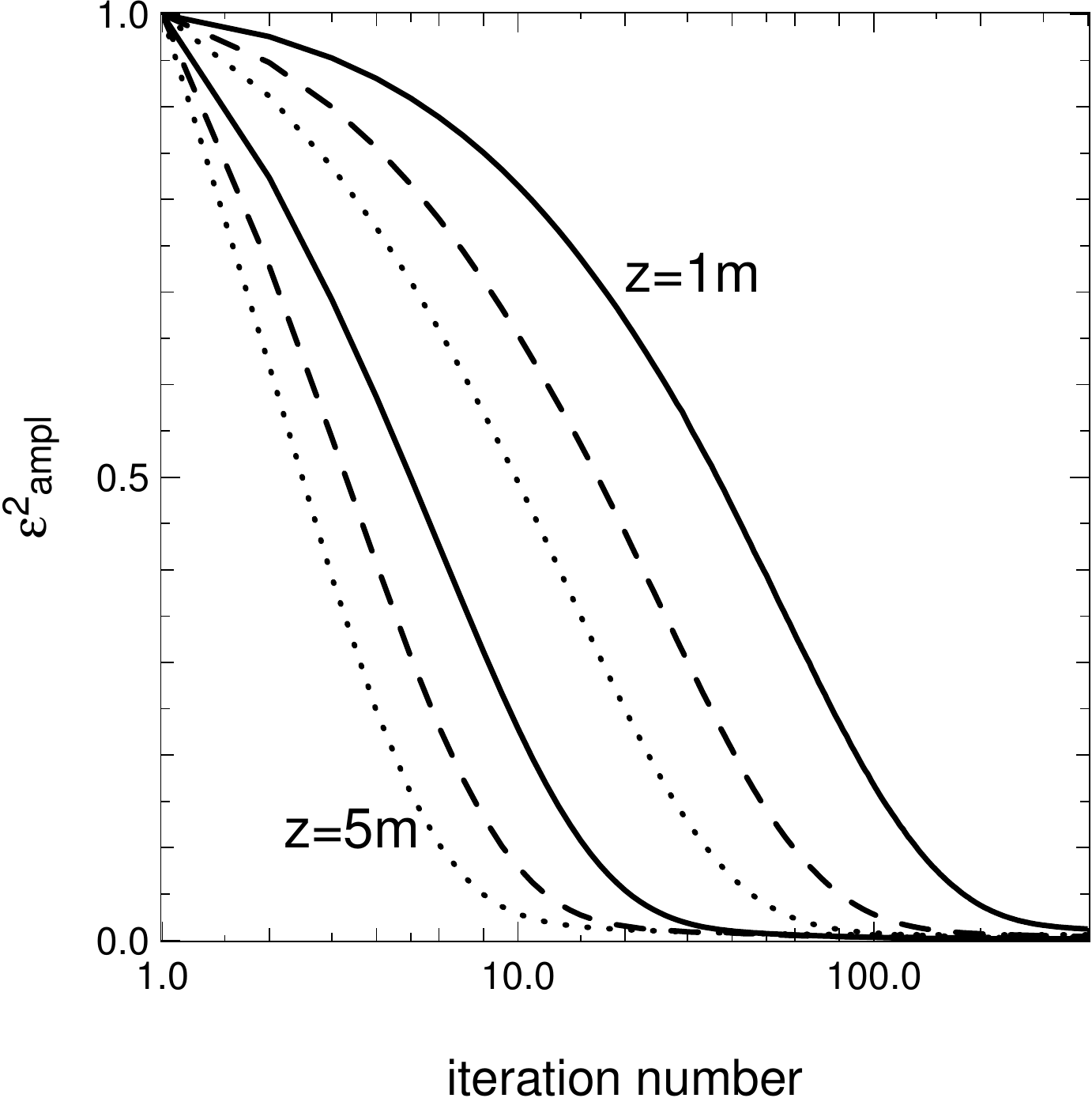}
  \hspace{0.5cm}
  \includegraphics[width=0.4\linewidth]{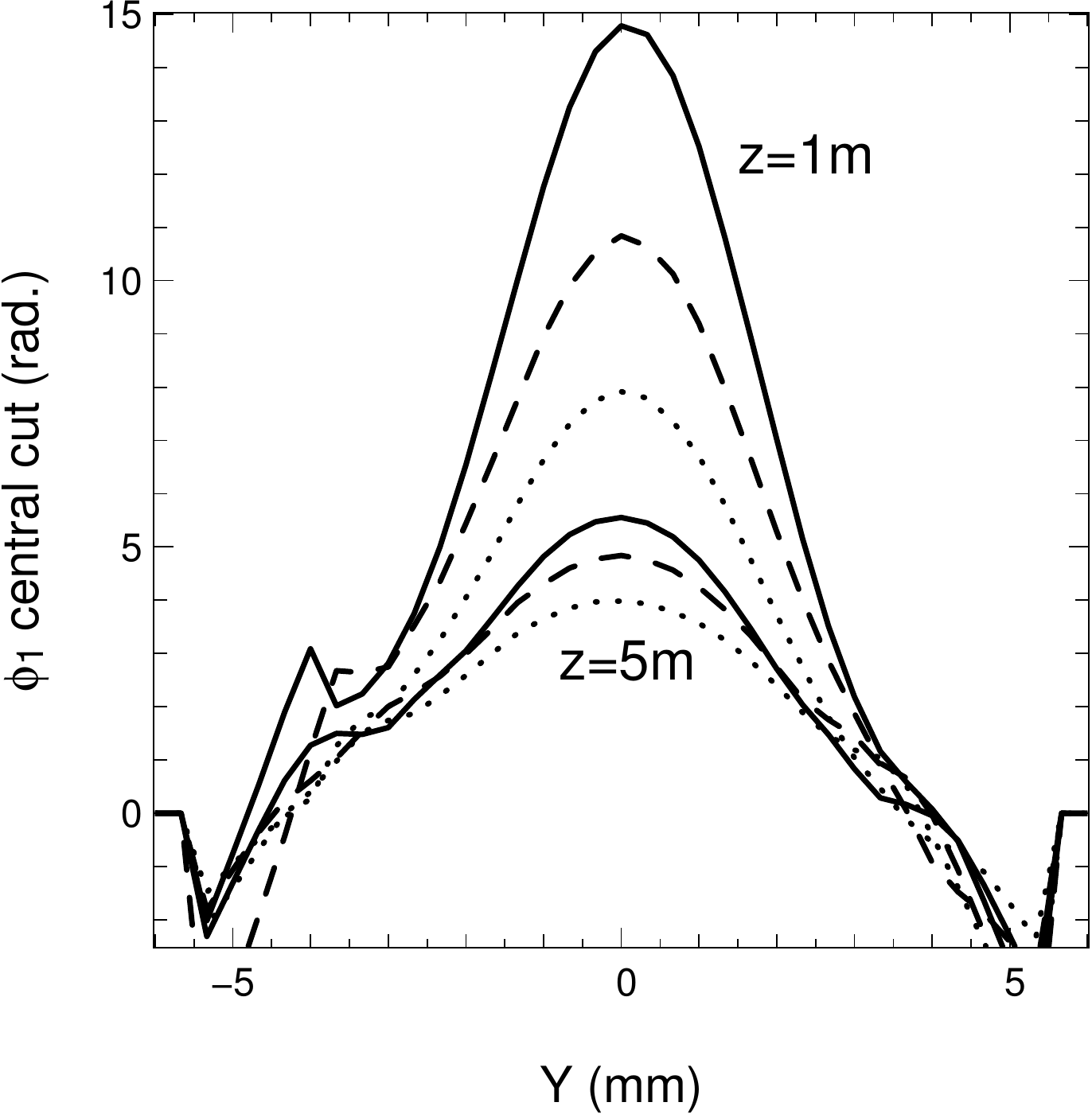}
  \caption{{\bf Left:} Evolution of $\epsilon^2_{\rm ampl}$ along with
    the iterations, for various DMs separations $z=1$, 1.5, 2, 3, 4
    and 5 meters. The slowest convergence is for $z=1~$m. {\bf Right:}
    Central cut along $y$ of the final $\phi_1$ deformation to be
    applied for the same DMs separations $z=1$, 1.5, 2, 3, 4 and 5
    meters. The largest phase amplitude is required for $z=1$~m and
    the smallest for $z=5~$m.}
\label{fig:test3-power}
\end{figure}

In addition to the evolution of the correction quality with the DMs
separation $z$, its influence on the convergence speed is also
studied. In Fig.~\ref{fig:test3-power} (left), the relative amplitude
error of the output beam $\epsilon^2_{\rm ampl}$ is represented as a
function of the iteration number. The slowest convergence is obtained
for the shortest distance $z=1$~m, where convergence is reached after
more than 350 iterations. The convergence is accelerated progressively
when the separation $z$ is increased. For $z=5~$m, less than 20
iterations suffice.

The convergence acceleration is due to the fact that a phase
deformation of minor amplitude is required when the separation
distance $z$ increases. This is illustrated on the right of
Fig.~\ref{fig:test3-power}, where the highest curve is the phase
deformation $\phi_1$ required for separation $z=1~$m. The required
deformation is reduced by approximately a factor equal to $p$ when the
separation is changed from $z=1~$m to $z=p$, with $p$ in the range of
1 to 5~meters. It is thus an important parameter to take into account
in the design of the beam shaping system, depending on the mechanical
properties of the DMs to be used.

\subsection{Influence of number of degrees of freedom}
\label{sec:infl-numb-dof}

In practice, expensive costs of the DMs with large
number of actuators may force to consider the application of phase
deformations with a lower spatial resolution than the one considered
in Sects.~\ref{sec:ampl-phase-corr} and \ref{sec:propag-z}. So a study
of the impact of the phase grid sampling for the estimates $\phi_1$
and $\phi_2$ on the performance of the beam shaping is required. We
specify this constraint in terms of the number of estimated phase
samples across the DMs diameter, $n$, such that a DM
with $\sim 3/4 n^2$ degrees of freedom over the pupil area is expected
to be able to reproduce it satisfactorily.

We choose the laser amplitude distortions $u_l$ as the ones measured
from GeMS in April and October 2013
(Fig.~\ref{fig:irrmap}). Again, we do not consider any phase
aberrations $\phi_l$ for the input. The simulation grid sampling has
been chosen equal to the resolution of the laser wavefront sensor
measurement at GeMS (Fig.~\ref{fig:irrmap}), \textit{i.e.}
$\delta=198~\mu$m. The DMs aperture diameter is asumed to be
4.9~mm. The desired output amplitude is set to a perfect Gaussian with
FWHM of 1.7~mm and a flat output phase.

\begin{figure}[htb!]
  \centering
  \includegraphics[width=0.4\linewidth]{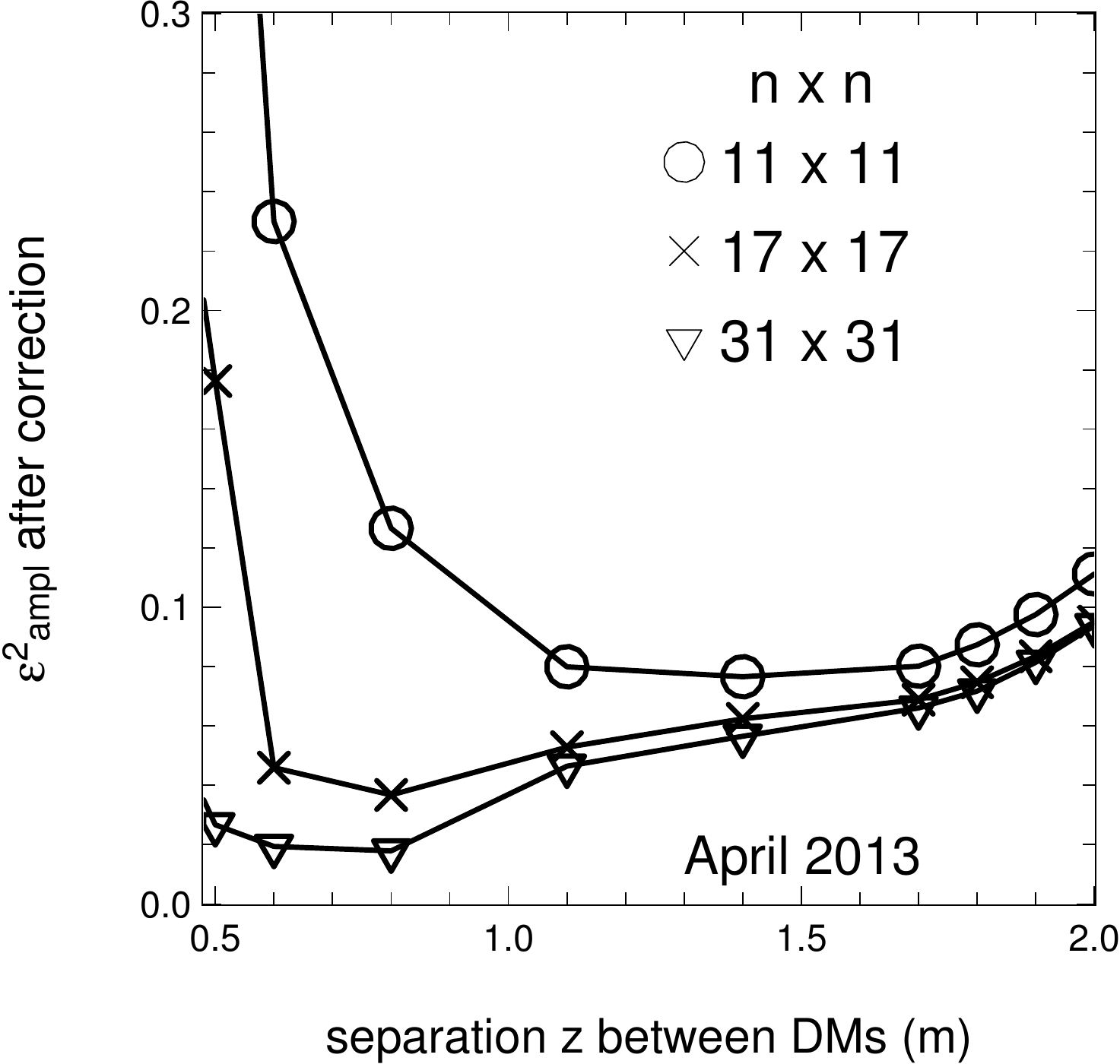}
  \hspace{0.5cm}
  \includegraphics[width=0.4\linewidth]{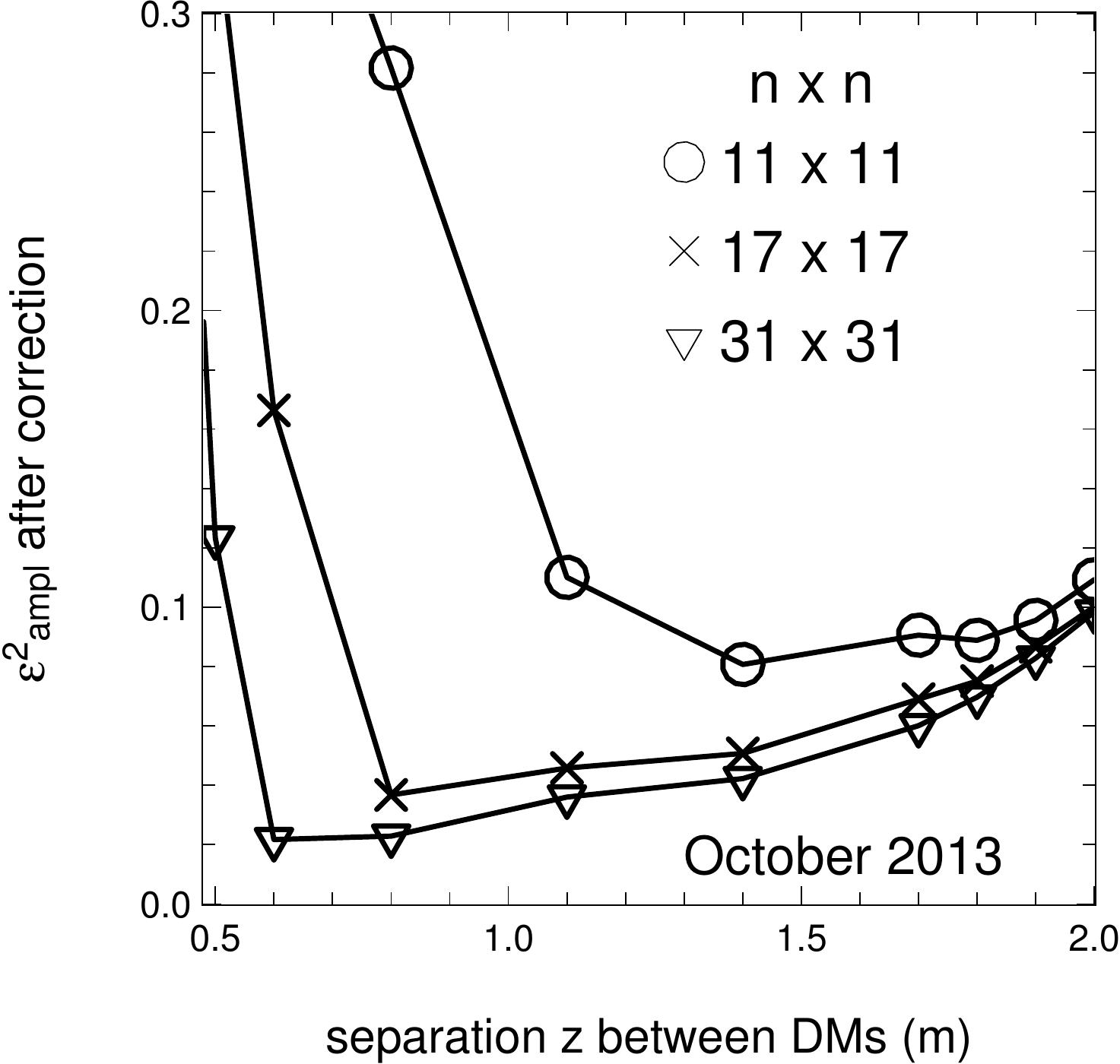}
  \caption{$\epsilon^2_{\rm ampl}$ after 400 iterations as a function
    of the separation $z$ between DMs, and for different number of
    samples $n$ across the pupil diameter. Amplitude distortions of
    input are taken from available measurements on GeMS laser in April
    2013 (left) and October 2013 (right). No phase distortions are
    assumed at input.}
\label{fig:ampl-error-test2-vs-n-varying-z-rel}
\end{figure}

The correction performance assessed by $\epsilon^2_{\rm ampl}$ as a
function of $z$ is represented in the plots of
Fig.~\ref{fig:ampl-error-test2-vs-n-varying-z-rel} for various values
of $n$. Input amplitude distortions measured in April 2013 are used
for the plot on the left and the ones measured in October 2013 are
used on the right plot. With a large number of degrees of freedom for
the correction, it is possible to reduce the error to a few percentage
points of its original value placing the DMs at short distances ($z$
below 1~m). At such short separations, the error reduction
significantly degrades when the number of phase samples $n$ across the
pupil is too small.

If the mirrors have only 11 actuators across the aperture for
instance, the remaining error is about 10\% of its original value, and
it requires greater separation distance $z$, \textit{i.e.} beyond
1.2~m.  Note that the distortions measured in October 2013 appear more
difficult to correct than the ones observed in April 2013. For
separations $z$ beyond 2 meters, there is no remarkable benefit in
using $31 \times 31$ phase samples compared to using $11 \times 11$.

\begin{figure}[htb!]
  \centering
  \includegraphics[width=0.4\linewidth]{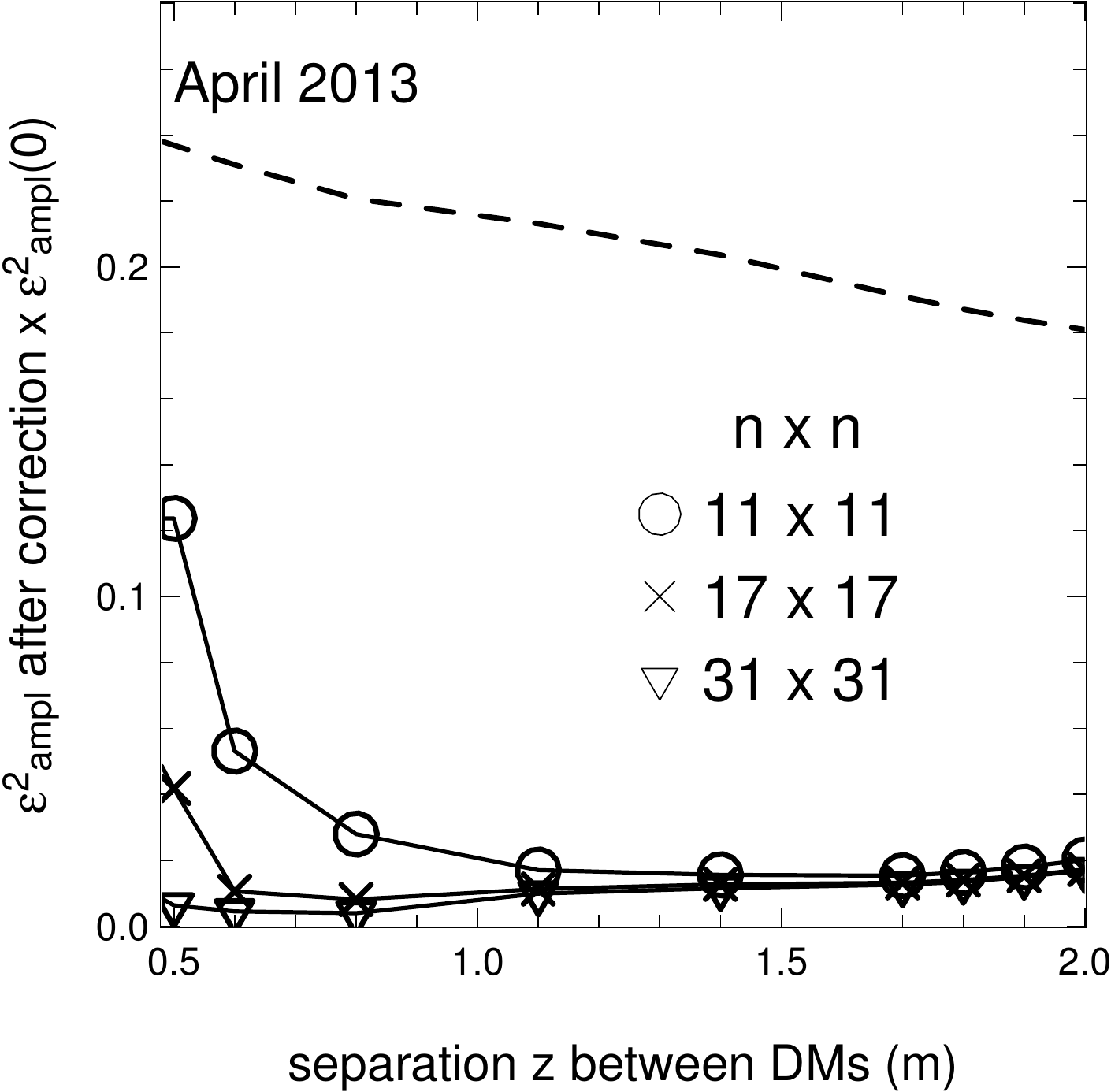}
  \hspace{0.5cm}
  \includegraphics[width=0.4\linewidth]{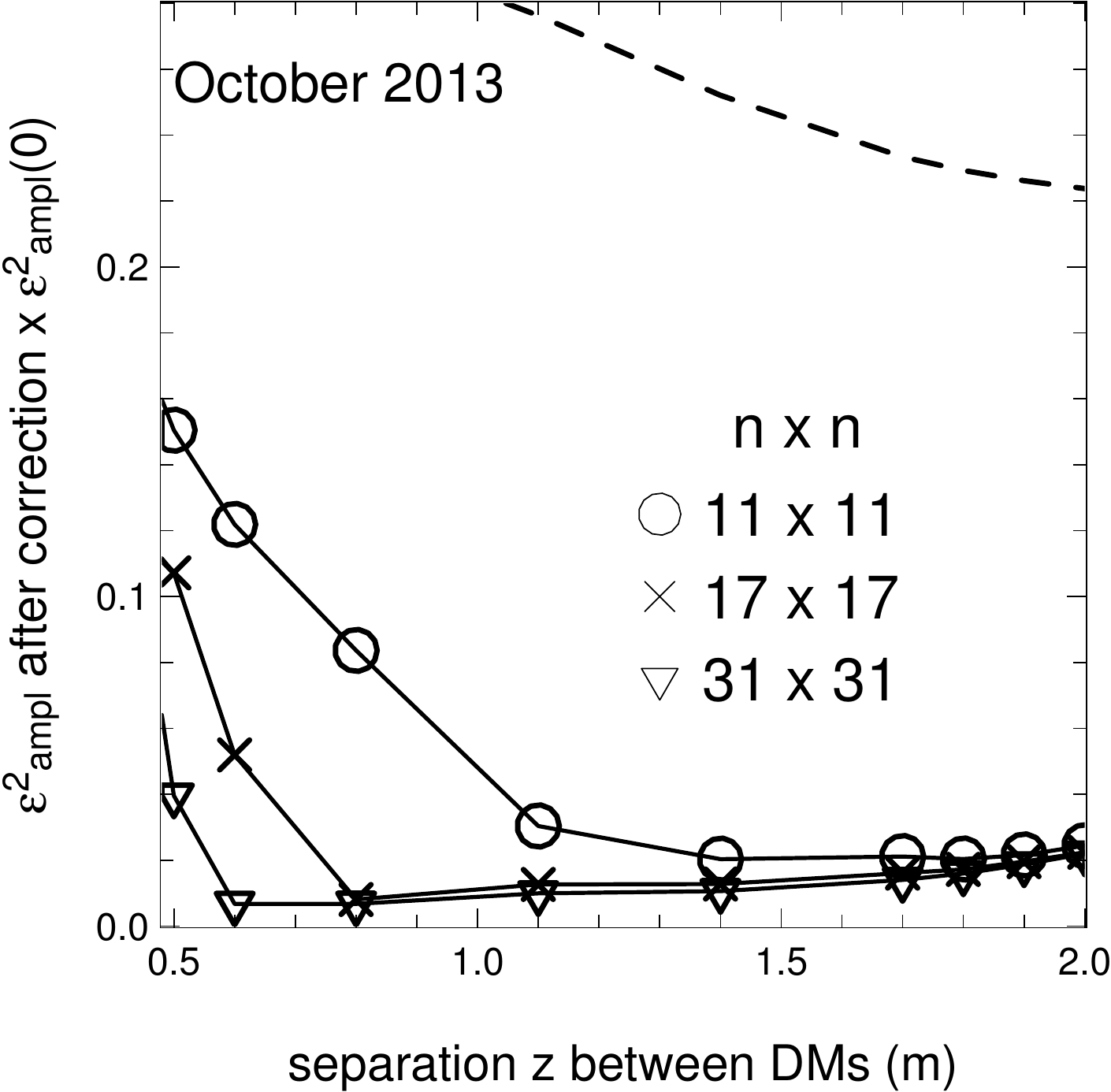}
  \caption{Same as in
    Fig.~\ref{fig:ampl-error-test2-vs-n-varying-z-rel} but the error
    is presented as an absolute value, not relatively to
    $\epsilon^2_{\rm ampl}(0)$. The dashed lines represent
    $\epsilon^2_{\rm ampl}(0)$ as a function of $z$.}
\label{fig:ampl-error-test2-vs-n-varying-z-abs}
\end{figure}

It is worth noting that a minimum for the relative error exists within
the DM separation range studied. For instance, if $n=31$, the relative
error $\epsilon^2_{\rm ampl}$ with $z=2~$m, is degraded by about 10
percentage points compared to the achieved error with
$z=80~$cm. However, this is mainly due to the normalization value in
the denominator of $\epsilon^2_{\rm ampl}$ (Eq.~(\ref{eq:crit-dist})),
which decreases with $z$. In order to clarify this, the same
simulations results are presented in
Fig.~\ref{fig:ampl-error-test2-vs-n-varying-z-abs} but showing the
amplitude error $\epsilon^2_{\rm ampl}$ before its normalization by
the error without correction in Eq.~(\ref{eq:crit-dist}). The
amplitude error without any beam shaping correction (dashed line)
progressively decreases with increasing separation distance $z$
between the DMs. This effect simply results from the propagation of
the beam in the near-field. This absolute representation reveals that
good correction quality is obtained if $z$ is greater than 1.4~m for
all three cases ($n=11$, 17 and 31). If $n$ is greater than 11,
shorter separations $z$ can be chosen for approximately the same good
correction quality.

\subsection{Benefit of correcting GeMS amplitude and phase distortions}
\label{sec:corr-gems}

\begin{figure}[htb!]
  \centering
  \includegraphics[width=0.4\linewidth]{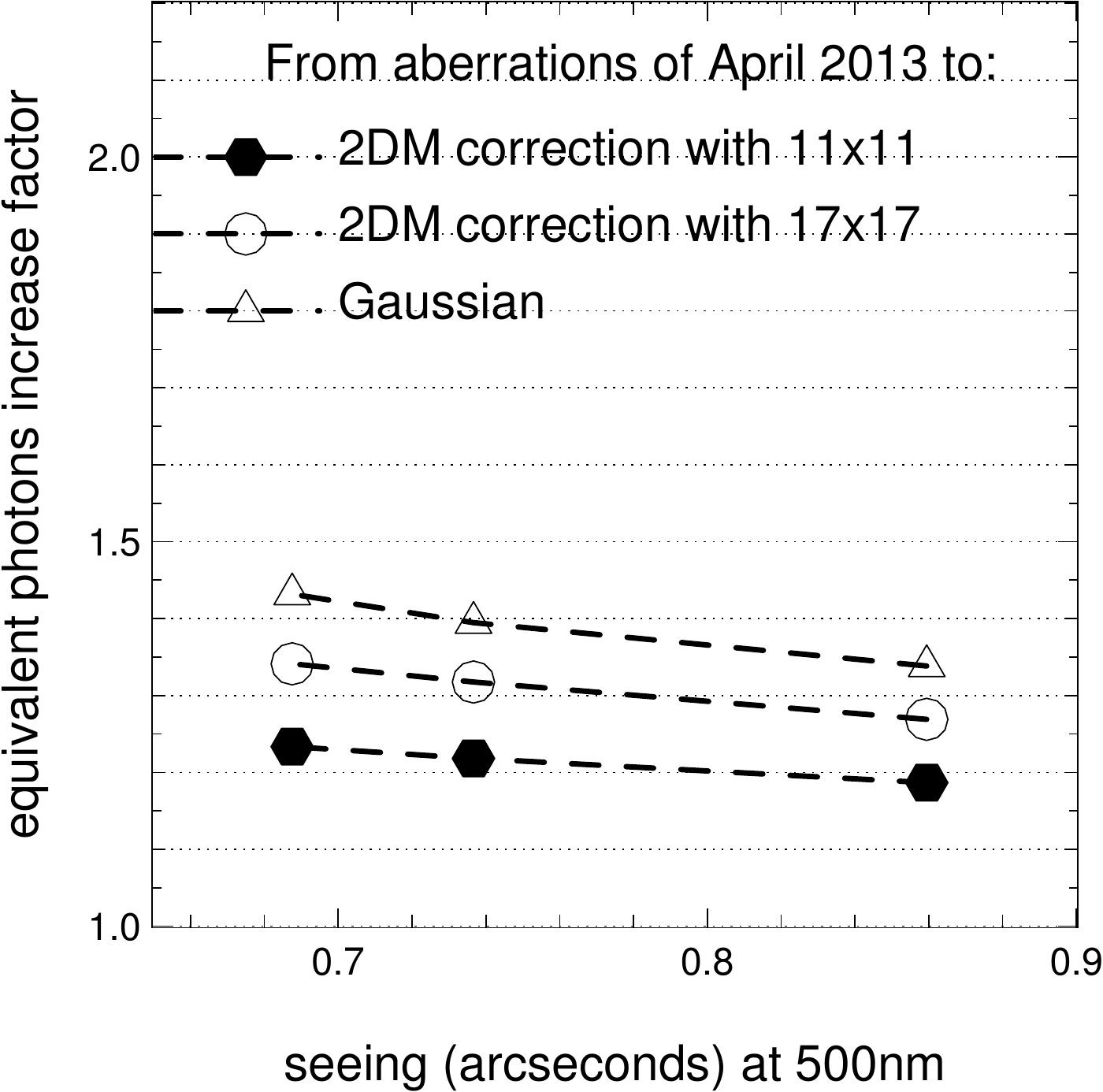}
  \hspace{0.5cm}
  \includegraphics[width=0.4\linewidth]{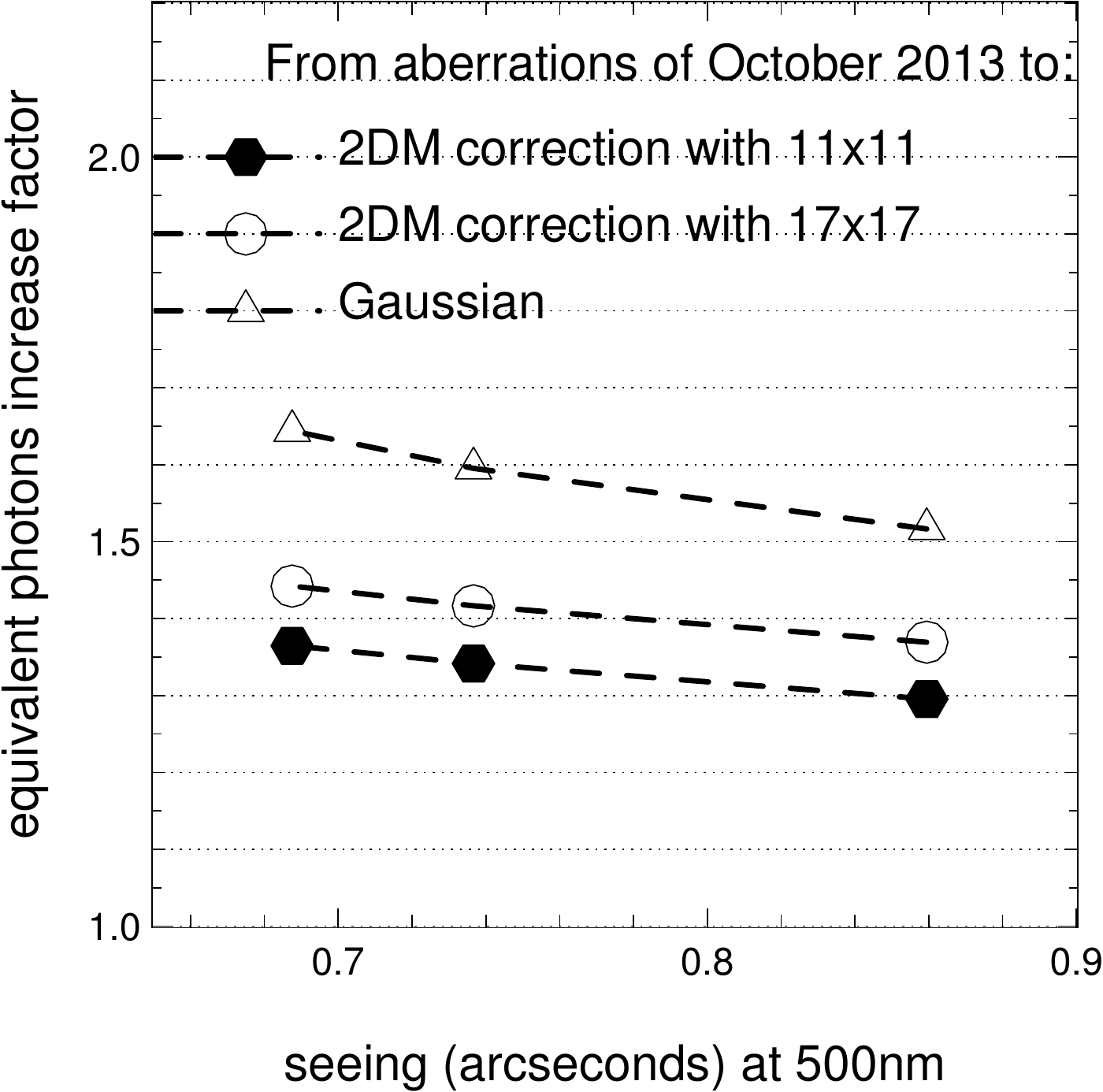}
  \caption{Equivalent increase of photons corresponding to the laser
    spot size reduction when launching $(i)$ a Gaussian beam, $(ii)$
    the beam corrected with $17 \times 17$ devices separated by 80~cm
    and $(iii)$ the beam corrected with $11 \times 11$ devices
    separated by 1.4~m. The reference laser spot size is the one
    produced by the laser aberrations measured at GeMS: {\bf Left:}
    aberrations of April 2013. {\bf Right:} aberrations of October
    2013.}
\label{fig:photonsfactor}
\end{figure}

According to the results of Sects.~\ref{sec:propag-z} and
\ref{sec:infl-numb-dof}, we are now interested in quantifying how much
the 2-DM correction system could reduce the size of the laser spot
image. For this, the measurements made at GeMS in April and October
2013 are considered with both their amplitude and wavefront
aberrations. The quality of the correction is evaluated by simulating
the uplink propagation of the resulting beam at the launching aperture
to the sodium layer as mentioned in Sect.~\ref{sec:GemsContext}. The
spot size at 90~km and its intensity are compared to the ones
evaluated without correction. The reduction of the laser spot size at
the sodium layer because of the correction is represented as an
equivalent increase of the number of photons (in
Eq.~(\ref{eq:sigmaE})) for a given AO noise level. These factors of
increase are shown in Fig.~\ref{fig:photonsfactor} as a function of
the simulated atmospheric seeing. Left graph stands for the
aberrations measured in April 2013, and right graph is for the ones of
October 2013. The reduction of the spot size, in terms
of photons increase, produced by a perfect Gaussian beam is also
plotted.

First, one observes that the reduction of the spot size is more
significant in good seeing conditions. The laser beam aberrations thus
degrade the AO performance more severely when the turbulence
conditions are good.

Second, because the 2-DM correction does not lead to a perfect
Gaussian beam the three curves are not superimposed with the upper one
(Gaussian). Still, with a correction system using $11\times 11$
mirrors, the equivalent increase of photons is in the range 20 to
40\%. A higher-order correction with $17$ actuators across the
aperture improves the photons increase to the range of 25 to
45\%. Obviously the correction of GeMS laser aberrations will not
allow to double the AO frame rate (which would require an equivalent
increase by 100\%), but still for median seeing of observing
conditions for GeMS an increase of the number of photons by 20 to 35\%
would already be achievable. In the most favorable conditions for a
correction with 11 actuators across the diameter, the frame rate could
even be increased by a factor 1.4, which could mean a jump from 350~Hz
to almost 500~Hz.

\section{Conclusion}
\label{sec:conclusion}

A two-deformable-mirror concept is developed to optimize the beam
shape of the laser used as beacons for adaptive optics in
astronomy. The optimization system aims at correcting quasi-static
beam distortions in amplitude and phase thanks to a phase retrieval
technique. Near-field algorithms previously developed in the
literature for defense and space communications applications are
adapted to the projection of lasers in the astronomical context. In
particular, an iterative and regularized unwrapping method is
developed to deduce commands for the mirrors. The method is
computationally efficient and robust to handle any input
amplitude distortions and any desired output amplitude, thanks to its
iterative feature and on-the-fly update of the weighting matrix
used for the unwrapping.

Preliminary simulations are presented to illustrate the effective
amplitude and phase correction achieved by the method. This correction
reduces the laser spot size thus improving the AO measurement
accuracy. The convergence is shown to be accelerated when the
separation distance between the 2 DMs increases. Regarding the number
of degrees of freedom required for the mirrors, a reasonable order of
11 actuators across the aperture is shown to provide good beam shaping
correction when applied to cases of amplitude and phase distortions
measured at GeMS during 2013. Additional degrees of freedom on the
deformable mirrors would improve the correction quality. Since the
algorithm converges in a few seconds, it can definitely be applied to
the the quasi-static aberrations observed at GeMS, which evolve
at the time-scale of hours or days.

The simulated cases highlight the fact that for reduced number of
degrees of freedom the considered examples of distortions are better
corrected with DMs separations greater than 1.4 meter approximately.
The simulation of uplink propagation toward the sodium layer show that
the correction of aberrations measured on GeMS laser will not allow to
double the AO frequency. Still in median seeing conditions for GeMS
observations a 2-DM correction system with 11 actuators across the
aperture may allow a reduction of the spot size by 10 to 15\%
compared to the spot generated by the laser beam with aberrations. For
a given AO noise level, this is equivalent to an increase of the
number of received photons by up to 40\%. Space constraints within the
telescope and costs of DMs will drive the trade-off between DM
separation and number of degrees of freedom when designing the
prototype to be tested on GeMS.

Finally, the measured quasi-static aberrations of the laser shown in
this paper are the ones observed after alignment and manual
optimization procedures on GeMS laser beam. This corresponds to a
time-consuming process usually involving 1 or 2 laser engineers for at
least 1 week before every run of GeMS. The presented 2-DM correction
can relax the need for such complex alignment optimizations. Actually
the laser optimization procedure at GeMS includes a trade-off between
beam quality and laser power. So relaxing in a first step the pressure
on laser beam quality could lead to improved laser power
\cite{FesquetAraujo2013a}.  

\section*{Aknowledgments}
\label{sec:aknowledgments}

This work was supported by the Chilean Research Council (CONICYT)
through grant Fondecyt 1120626. Benoit Neichel aknowledges the French
ANR program WASABI - ANR-13-PDOC-0006-01. Particular thoughts and
aknowledgements go to our co-author Vincent Fesquet who recently
passed away.

\end{document}